\documentclass[preprint,12pt]{elsarticle}

\usepackage{amssymb}
\usepackage{amsmath}
\usepackage{amsfonts}
\usepackage[]{algorithm2e}
\usepackage{caption}
\usepackage{subcaption}
\usepackage{marginnote}
\usepackage{color}
\usepackage{lineno}
\usepackage{natbib}
\journal{Journal of Computational Physics}

\begin{document}

\begin{frontmatter}

%% Title, authors and addresses

%% use the tnoteref command within \title for footnotes;
%% use the tnotetext command for theassociated footnote;
%% use the fnref command within \author or \address for footnotes;
%% use the fntext command for theassociated footnote;
%% use the corref command within \author for corresponding author footnotes;
%% use the cortext command for theassociated footnote;
%% use the ead command for the email address,
%% and the form \ead[url] for the home page:
%% \title{Title\tnoteref{label1}}
%% \tnotetext[label1]{}
%% \author{Name\corref{cor1}\fnref{label2}}
%% \ead{email address}
%% \ead[url]{home page}
%% \fntext[label2]{}
%% \cortext[cor1]{}
%% \address{Address\fnref{label3}}de
%% \fntext[label3]{}

\title{Sequential Fully Implicit Formulation for Compositional Simulation using Natural Variables}

\author[labelTOTAL,labelETH]{A. Moncorg\'e \corref{cor1} }
\author[labelStanford]{H.A. Tchelepi}
\author[labelETH]{P. Jenny}

\address[labelTOTAL]{Geoscience Research Center, TOTAL E\&P UK, Aberdeen, UK.}
\address[labelETH]{Institute of Fluid Dynamics, ETH Zurich, Zurich, Switzerland.}
\address[labelStanford]{Energy Resources Engineering Department, Stanford University, Stanford, California.}
\cortext[cor1]{Corresponding author arthur.moncorge@total.com}

\begin{abstract}

The Sequential Fully Implicit (SFI) method was proposed (Jenny et al., JCP 2006), in the context of a Multiscale Finite Volume (MSFV) formulation, to simulate coupled immiscible multiphase fluid flow in porous media.
Later, Lee et al. (Comp. Geosci. 2008) extended the SFI formulation to the black-oil model, whereby the gas component is allowed to dissolve in the oil phase. Most recently, the SFI approach was extended to fully compositional isothermal displacements by  Moncorg\'e et al., (JCP 2017).
SFI schemes solve the fully coupled system in two steps:
(1) Construct and solve the pressure equation (flow problem). (2) Solve the coupled species transport equations for the phase saturations and phase compositions. In SFI, each outer iteration involves this two-step sequence.
Experience indicates that complex interphase mass transfer behaviors often lead to large numbers of SFI outer iterations compared with the Fully Implicit (FI) method. 
Here, we demonstrate that the convergence difficulties are directly related to the treatment of the coupling between the flow and transport
problems, and we propose a new SFI variant based on a nonlinear overall-volume balance equation. The first step consists of forming and solving a nonlinear pressure equation, which is a weighted sum of all the component mass conservation equations. A Newton-based scheme is used to iterate out all the pressure dependent 
nonlinearities in both the accumulation and flux terms of the overall-volume balance equation. The resulting pressure field is used to compute the Darcy phase velocities and the total-velocity. 
The second step of the new SFI scheme entails introducing the overall-mass density as a degree-of-freedom, and solving the full set of component conservation equations cast in
the natural-variables form (i.e., saturations and phase compositions). During the second step, the pressure and the total-velocity fields are fixed.
The SFI scheme with a nonlinear pressure extends the SFI approach of Jenny et al. (JCP 2006) to multi-component compositional processes with interphase mass transfer.  
The proposed compositional SFI approach employs an overall balance for the pressure equation; however, unlike existing volume-balance Sequential Implicit (SI) schemes 
(Acs et al., SPEJ 1985; Watts, SPEJ 1986; Trangenstein \& Bell, SIAM 1989; Pau et al., Comp. Geosci. 2012; Faigle et al., Comp. Meth. App. Mech. Eng. 2014; and Doster et al., CRC 2014), 
which use overall compositions, this SFI formulation is well suited for the natural variables (saturations and phase compositions). 
We analyze the `splitting errors' associated with the compositional SFI scheme, and we show how to control these errors in order to converge to the same solution as the Fully Implicit (FI) method.
We then demonstrate that the compositional SFI has convergence properties that are very comparable to those of the FI approach.
This robust sequential-implicit solution scheme allows for designing numerical methods and linear solvers that are optimized for the sub-problems of flow and transport.
The SFI scheme with a nonlinear pressure formulation is well suited for multiscale formulations, and it promises to replace the widely used FI approach for compositional reservoir simulation.

\end{abstract}

%% \fntext[label3]{}

%
\begin{keyword}
%% keywords here, in the form: keyword \sep keyword
nonlinear dynamics  \sep numerical flow simulation \sep sequential implicit  \sep operator splitting \sep coupled flow and transport \sep multiscale methods  \sep flow in porous media \sep compositional formulation 
\sep multiphase flow \sep multi-component transport \sep compositional reservoir simulation

%% PACS codes here, in the form: \PACS code \sep code

%% MSC codes here, in the form: \MSC code \sep code
%% or \MSC[2008] code \sep code (2000 is the default)

\end{keyword}

\end{frontmatter}

%% \linenumbers

%% main text

\tableofcontents

\section{Introduction}

The Sequential Fully Implicit (SFI) method was first proposed to model multiphase fluid flow without mass exchange in the context of the 
Multiscale Finite Volume (MSFV) method~\cite{Jenny:2006}.
The SFI approach was later extended to the black-oil formulation (three pseudo components with gas dissolved in the oil phase) \cite{Lee:2008,MoynerBO:2016}, and recently 
for compositional multi-component flow by Hajibeygi \& Tchelepi \cite{Hajibeygi:2014}, Moncorg\'e et al. \cite{Moncorge:2016,Moncorge:2017} and M{\o}yner et al. \cite{MoynerRSC:2017,MoynerIPAM:2017}.
In these previous works, the SFI algorithm consists of solving two different implicit systems in sequence: 
(1) a pressure equation, and (2) a saturation/composition system.
During the solution of the saturation/composition system, the pressure is fixed. 
%it is equivalent to assuming that the total material balance is solved.
For $n_c$ components, a coupled set of $n_c -1$ component conservation equations is solved in the saturation/composition step. 
Our objective is to design an SFI scheme with a nonlinear convergence rate that is comparable to that of the Fully Implicit (FI) formulation across
the full range of compositional modeling of practical interest. This objective is motivated by several factors. First, multiscale formulations rely on sequential
treatment of the flow (near-elliptic) and transport (hyperbolic) problems; thus, in order for multiscale methods to fully replace
the existing single-level formulations, the SFI approach must have nonlinear convergence rates that are comparable, or even superior,
to Fully Implicit (FI) solution schemes. Second, sequential treatment of flow and transport makes it possible to employ advanced discetization schemes and scalable 
solution methods to the different equations in the coupled system. 

Moncorg\'e et al. \cite{Moncorge:2017} enriched the pressure equation with Fully-Implicit (FI) coupling in regions that experience phase appearance
to ensure convergence of the full system to any given tolerance. 
This approach is efficient as long as the fraction of cells that have both liquid and vapor phases is relatively small.
However, as the number of cells that experience phase appearance/disappearance grows, this approach becomes expensive.
In Lee at al.~\cite{Lee:2008} and M{\o}yner et al.~\cite{MoynerBO:2016,MoynerRSC:2017,MoynerIPAM:2017}, 
the conservation equation of one component is removed
when solving the saturation/composition system.
Hajibeygi \& Tchelepi \cite{Hajibeygi:2014} tried to remove the material-balance error by freezing the overall-mass density in the accumulation term 
and the total-mass flux between the computational cells. 
Their algorithm is convergent; however, in cases with strong coupling, the scheme needs large numbers of outer iterations. 
Acs et al. \cite{Acs:1985}, Watts \cite{Watts:1986}, Trangenstein \& Bell \cite{TBBO:1989,TBC:1989}, Dicks \cite{Dicks:1993},
Pau et al. \cite{Pau:2012}, Faigle et al. \cite{Faigle:2014} and Doster et al. \cite{Doster:2014}
developed Sequential Implicit (SI) formulations for compositional multi-component displacements.
In these SI formulations, extensive degrees-of-freedom (variables), such as the overall number of moles, or mass, of a component divided by the pore volume, are used.

The developers of SI formulations noted that it is not possible to satisfy all the governing nonlinear equations exactly, 
and that some inconsistencies must be tolerated.
The formulations of Acs et al., Watts, Trangenstein \& Bell, Dicks, Pau et al., Faigle et al. and Doster et al. consider the pressure equation as a `volume balance' equation;  
in the second step of solving the transport, they keep the conservation equations of all the components.
Upon convergence, the mass conservation equation of each of the components is satisfied subject to the desired tolerance; 
however, some discrepancies in the overall-volume balance and the total-velocity persist.
Only the overall-volume balance splitting error has been defined in these previous works.
Attempts to reduce this volume splitting error have been done in Acs et al., Trangenstein \& Bell, Pau et al., Faigle et al. and Doster et al. 
by using a local relaxation term to keep the error bounded in time or by local smoothing by Dicks. 
They all result in local changes relaxing the mass balance equations.
However, these volume splitting errors are very local and have rarely large effect on the overall flow.
On the contrary, the total-velocity splitting error, has never been documented before and has a much larger support.
We show that we need to control this second splitting error to recover the fully-implicit solution.

Acs et al., Watts, Trangenstein \& Bell, Pau et al., Faigle et al. and Doster et al. employ a linearized `volume balance' equation.
Coats \cite{Coats:2000} proposed an IMPES (IMplicit Pressure Explicit Saturations) pressure equation built by algebraic manipulations of the usual material balance equations.
M{\o}yner et al. \cite{MoynerIPAM:2017,MoynerSIAM:2017} use this pressure for their sequential implicit method for compositional flow.
The derivatives of the overall compressibility wrt pressure and the derivatives of the partial molar volumes wrt pressure are not accounted for in the linearized `volume balance' equation 
and the derivatives of the partial molar volumes wrt pressure are not accounted for in the Coats pressure equation. 
Both equations experience nonlinear convergence difficulties, especially when the coupling between the multiphase flow and the multi-component transport is strong. 
Here, we propose a nonlinear `volume balance' equation that accounts for all the necessary derivatives wrt pressure.
For immiscible and incompressible multiphase flow, the nonlinear pressure equation that we derive here simplifies to the pressure equation of Acs et al., Watts, Trangenstein \& Bell, Pau et al., Faigle et al. Doster et al.
as well as the pressure equations of Coats and Young \cite{Young:2001}. 
It is also identical to the pressure equation for immiscible multiphase flow of Jenny et al. \cite{Jenny:2006} 
and to the pressure equation of Lee et al. \cite{Lee:2008} and M{\o}yner \& Lie \cite{MoynerBO:2016} for the black-oil formulation.

This new SFI scheme with a nonlinear pressure equation can be seen as an adaptation of existing SI methods based on overall compositions to one suited for the natural-variables formulation.
The splitting of the coupled system into  two nonlinear problems, namely, a nonlinear pressure equation and a nonlinear multi-component transport problem, leads to volume-balance and total-velocity errors.
We analyze these `splitting errors' , and we show how to control them.
We then demonstrate that the parabolic pressure equation and the hyperbolic transport systems are
easier to solve separately using widely available linear solvers compared with the specialized solvers (e.g., CPR-AMG \cite{CPR:2005}) needed to
deal with the full system with mixed parabolic-hyperbolic behaviors.

The paper is organized as follows. 
First, we introduce a natural-variables formulation, which
entails forming a nonlinear volume-balance equation followed by solving the full set of transport equations.
Then, we elaborate the details of constructing the volume-balance equation.
We also analyze the splitting errors and how to control them.
We finally demonstrate the performance of the compositional SFI scheme for very challenging test cases.

%% main text
\section{Sequential Fully Implicit Scheme for Compositional Flow}

In this section, we describe the compositional formulation based on natural-variables for both Fully Implicit (FI) and Sequential Fully Implicit (SFI) nonlinear 
solution algorithms.
We present a variant of the natural-variables formulation, whereby the system is augmented by an additional equation and an additional variable.
Following that, the new SFI scheme for compositional flow is described.

\subsection{Compositional Formulation Based on Extended Natural-Variables}
%%%
The generalized compositional formulation, which accounts for multi-component, three-phase flow with interphase mass transfer, is the target of this work. 
%%%
We assume that we have $n_c$ components, i.e., $n_h$ hydrocarbon components and the water component, and we employ the natural-variables formulation \cite{Coats:1980}. 
The overall density, $\rho_t$, is used as an additional global variable.
The number of unknowns in the case of three-phase flow is $2n_h+7$. They are: gas-phase pressure, $P_g$, oil-phase pressure, $P_o$, water-phase pressure, $P_w$, total mole density ${\rho}_t$, 
the saturations of each phase, $S_g$, $S_o$, $S_w$, the component mole (mass) fractions in the gas, ${y}_c$, and the liquid, ${x}_c$.
%%%
To close the system, $2n_h+7$ equations are required. 

\subsubsection{Mass conservation equations}

The conservation of a hydrocarbon component, $c$, and of the water component, $w$, can be written as:
%%%
\begin{eqnarray}
\frac{\partial }{\partial t} \left[ \phi \left( {y}_c{\rho}_g S_g +{x}_c{\rho}_o S_o \right) \right] + \nabla \cdot \left[{y}_c{\rho}_g u_g +{x}_c{\rho}_o u_o  \right] ={y}_c{\rho}_g q_g +{x}_c{\rho}_o q_o \label{M_c}\\
\mbox{and } \frac{\partial }{\partial t} \left[ \phi {\rho}_w S_w \right] + \nabla \cdot \left[{\rho}_w u_w \right] = {\rho}_w q_w  \label{M_w},
\end{eqnarray}
where $\phi$ is the porosity, $\rho_g$, $\rho_o$ and $\rho_w$ are the fluid-phase mole densities, and $q_g$, $q_o$ and  $q_w$ source terms.
The velocity of each phase $p\in\{g,o,w\}$ is given by Darcy's law:
\begin{equation}
u_p = -\frac{k_{r_p}}{\mu_p} K \left( \nabla P_p - \overline{\rho}_p g \nabla D \right) = -\lambda_p K \left( \nabla P_p - \overline{\rho}_p g \nabla D \right),
\label{UP_DARCY}
\end{equation}
where $K$ is the rock permeability, $k_{r_p}$ the relative permeability of each phase, $\mu_p$ the phase viscosities, $\lambda_p=\frac{k_{r_p}}{\mu_p}$ the phase mobilities, $\overline{\rho}_p$ the fluid-phase mass densities, $g$ the gravitational acceleration, $D$ the depth.
The total-velocity 
\begin{equation}
u_t = - \sum_p \lambda_p K \left( \nabla P_p - \overline{\rho}_p g \nabla D \right)
\label{UT}
\end{equation}
 is the sum of the phase velocities. The phase velocities
\begin{equation}
u_p = \frac{\lambda_p}{\lambda_t}  u_t + \sum_q \frac{\lambda_p \lambda_q}{\lambda_t}  K \left[ \left( \nabla P_q - \overline{\rho}_q g \nabla D \right) - \left( \nabla P_p - \overline{\rho}_p g \nabla D \right) \right]  = \frac{\lambda_p}{\lambda_t} u_t + \Psi_p
\label{UP_UT}
\end{equation}
can be expressed as a function of the total-velocity,
where $\lambda_t=\sum_p \lambda_p$ is the total-mobility and
$\Psi_p = \sum_q \frac{\lambda_p \lambda_q}{\lambda_t} K \left[ \left( \nabla P_q - \overline{\rho}_q g \nabla D \right) - \left( \nabla P_p - \overline{\rho}_p g \nabla D \right) \right]$ represents the capillary pressure and gravity contributions for each phase. 
Writing the phase velocities in terms of the total-velocity  (\ref{UP_UT}), and substituting into equations (\ref{M_c}) and (\ref{M_w}), rewriting the source terms, 
and adding the total-mole density ${\rho}_t$ as an additional variable in the accumulation term 
leads to:

\begin{eqnarray}
\frac{\partial }{\partial t} \left[ \phi \frac{{y}_c{\rho}_g S_g +{x}_c{\rho}_o S_o}{ \sum_p {\rho}_p S_p} {\rho}_t \right] + \nabla \cdot \left[{y}_c{\rho}_g  \left( \frac{\lambda_g}{\lambda_t} u_t + \Psi_g \right) +{x}_c{\rho}_o  \left( \frac{\lambda_o}{\lambda_t} u_t + \Psi_o \right)  \right] \nonumber \\
= \left( {y}_c{\rho}_g \frac{\lambda_g}{\lambda_t} + {x}_c{\rho}_o \frac{\lambda_o}{\lambda_t} \right) q_t \label{M_u_c}\\
\mbox{and } \frac{\partial }{\partial t} \left[ \phi \frac{{\rho}_w S_w}{ \sum_p {\rho}_p S_p} {\rho}_t \right] + \nabla \cdot \left[ {\rho}_w  \left(\frac{\lambda_w}{\lambda_t} u_t+\Psi_w \right)  \right] = {\rho}_w \frac{\lambda_w}{\lambda_t} q_t, \label{M_u_w}
\end{eqnarray}
where $q_t=\sum_p q_p$. Equations (\ref{M_u_c}) and (\ref{M_u_w}) are the conservation equations in `transport form'.

\subsubsection{Constraint equations}

The $n_h+5$ constraint equations involve variables 
in the control-volume (cell) under consideration. We have:
\begin{eqnarray}
\sum_{c=1}^{n_h}{y}_c&=&1, \label{SUMY} \\
\sum_{c=1}^{n_h}{x}_c&=&1 \label{SUMX} \\
\mbox{and } S_g + S_o+ S_w&=&1.\label{SUMS}
\end{eqnarray}
%
%
% HAT-old: do we need Pc in this paper.
We also have the capillary-pressure versus saturation relationships:
\begin{eqnarray}
P_{c_{GO}}(S_g) = P_g - P_o\\
\mbox{and } P_{c_{OW}}(S_w) = P_o - P_w,
\end{eqnarray}
where $P_{c_{GO}}$ and $P_{c_{OW}}$ are the capillary pressures between gas and oil phases and the oil and aqueous phases, respectively.
The remaining constraints represent thermodynamic phase equilibrium for each hydrocarbon component ($n_h$ equations).
These local equilibrium constraints are applied only when both hydrocarbon phases (oil and gas) are present in the control volume. 
Existing formulations can be based on black-oil, or K-value correlations \cite{Crookston:1979}, as well as the Peng-Robinson \cite{PR:1976}, Redlich-Kwong \cite{RK:1949} and Soave-Redlich-Kwong \cite{SRK:1972} cubic equations of state (EOS) models. They can all be written as \cite{Martin:1979}:
\begin{equation}
\hat{f}_{c,g}(P_g,y_1,...,y_{n_h})= \hat{f}_{c,o}(P_g,x_1,...,x_{n_h}).
\label{FUGACITY}
\end{equation}

\subsubsection{Thermodynamic volume equation}

The mole-fraction of each phase is denoted $\beta_p(P,z)$ with $p\in\{g,o,w\}$. The overall mole-fraction of a component, $z := (z_\alpha)_{\alpha=c,w}$ can be
expressed in terms of the natural variables as:
\begin{eqnarray}
z_c &=& \frac{{y}_c{\rho}_g S_g +{x}_c{\rho}_o S_o}{ \sum_p {\rho}_p S_p} \label{Z_C}\\
\mbox{and } z_w &=& \frac{{\rho}_w S_w}{ \sum_p {\rho}_p S_p} \label{Z_W}.
\end{eqnarray}
The phase-split algorithm is used to solve the system of $2n_h+3$ equations 
and $2n_h+3$ variables $y_c$, $x_c$ and $\beta_p$ with the pressure, $P$, and the overall mole-fractions, $z$, fixed.
\begin{eqnarray}
z_c &=& {y}_c \, \beta_g + {x}_c \, \beta_o \label{PHASE_SPLIT_SYSTEM_ZC} \\
z_w &=& \beta_w \label{PHASE_SPLIT_SYSTEM_ZW} \\
\hat{f}_{c,g}(P_g,y_1,...,y_{n_h}) &=& \hat{f}_{c,o}(P_g,x_1,...,x_{n_h}) \label{PHASE_SPLIT_SYSTEM_FC} \\
\sum_p \beta_p &=& 1.0 \label{PHASE_SPLIT_SYSTEM_SUMB}  \\
\mbox{and } \sum_c y_c &=& 1.0. \label{PHASE_SPLIT_SYSTEM_SUMY}
\end{eqnarray}
Since $\sum_p \beta_p = 1.0$ and $\sum_{\alpha=c,w} z_{\alpha} = 1.0$, the last equation is equivalent to $\sum_c x_c = 1.0$.
To compute the thermodynamic values of the saturations and the mole-fractions, we first compute the overall mole-fractions
with (\ref{Z_C}) and (\ref{Z_W}), we then solve the phase-split system to get the thermodynamic values of the mole-fractions $y_c$ and $x_c$.
Finally, we use the $\beta_p$ to compute the
% 
% HAT: why normalized. Explain
normalized thermodynamic values of the saturations (normalized values needed for the relative permeabilities and the capillary pressures functions):
\begin{equation}
\overline{S}_p^{thermo} =  \frac{ \frac{\beta_p}{\rho_p} }{ \sum_q \frac{\beta_q}{\rho_q} }.
\label{THERMO_SATURATION_NORMALIZED}
\end{equation}
The phase-split computations follow the work in \cite{MichelsenMollerup:2007}.

In addition to the $n_c=n_h+1$ conservation equations and the $n_h+5$ constraint equations, we employ an additional constraint equation and an additional variable.
We choose the total-mole density, $\rho_t$, as the additional variable for two reasons.
First, $\rho_t$ is an intensive variable, but unlike saturation or a mole-fraction, it is not a fraction;
second, the phase-split computations are independent of $\rho_t$.
Let us define 
\begin{equation}
S_p^{thermo} =  \rho_t \frac{\beta_p}{\rho_p}
\label{THERMO_SATURATION}
\end{equation}
as the `thermodynamic saturation' of phase $p$. It is the volume of phase $p$ divided by the pore-volume.
At convergence, the sum of the thermodynamic volumes of the fluid phases must equal the pore volume.  

The usual phase saturation, $S_p$, is referred to as the `flow saturation'.
With the definition of the thermodynamic saturations, the following relationship is always satisfied:
\begin{equation}
\sum_p \rho_p S_p^{thermo} =  \sum_p \rho_p \left(\rho_t \frac{\beta_p}{\rho_p} \right) = \rho_t \sum_p \beta_p = \rho_t.
\label{TOTAL_MOLE_DENSITY}
\end{equation}
The additional constraint equation is the `thermodynamic volume equation', namely,
\begin{equation}
\sum_p S_p^{thermo} = \rho_t \sum_p \frac{\beta_p}{\rho_p} = 1.0.
\label{VOLUME_CONSTRAINT}
\end{equation}
If the thermodynamic volume equation is satisfied, then the thermodynamic saturations and the flow saturations are equal, and 
the following relationship holds:
\begin{equation}
\rho_t := \sum_p \rho_p S_p^{thermo} = \sum_p \rho_p S_p.
\label{VOLUME_CONSTRAINT_RHOT}
\end{equation}

\subsubsection{Primary and secondary equations and variables}
We have a nonlinear system of $2n_h+7$ equations and $2n_h+7$ unknowns.
The $n_c = n_h+1$ conservation equations are used as the primary equations. 
In addition to the `thermodynamic volume equation',
we have $n_h+5$ local constraint equations, which make up the secondary equations.
The secondary equations are local constraints and only involve the variables of the control-volume (cell) under consideration. 
When we linearize the system of equations, the secondary equations are used to eliminate algebraically the $n_h+5$ variables.
After the linear solve that computes the update of the primary variables, the update of the secondary variables are computed
from the linearization of the secondary equations.

In our formulation, the gas-phase pressure, $P_g$, is always a primary variable.
The remaining $n_h$ primary variables depend on the phase state of the control volume.
For oil-water cells, the primary variables are: $S_w$ and $x_1 \dots x_{n_h-1}$;
for gas-water cells, the primary variable set is: $S_w$ and $y_1 \dots y_{n_h-1}$; 
for three-phase cells, the variables are: $S_g$, $S_w$ and $y_2 \dots y_{n_h-1}$.

The sum of all the component conservation equations leads to:
\begin{equation}
\frac{\partial }{\partial t} \left[ \phi \sum_p {\rho}_p S_p \right] + \nabla \cdot \left[ \sum_p \rho_p u_p  \right] = \sum_p \rho_p q_p. 
\label{M_t}
\end{equation}
Substitution of Eq.\ (\ref{VOLUME_CONSTRAINT_RHOT}) into the accumulation term and Eq.\ (\ref{UP_UT}) in the flux term leads to
\begin{equation}
\frac{\partial }{\partial t} \left[ \phi \rho_t \right] + \nabla \cdot \left[ \sum_p {\rho}_p  \left( \frac{\lambda_p}{\lambda_t} u_t+\Psi_p \right)  \right] = \sum_p {\rho}_p \frac{\lambda_p}{\lambda_t} q_t. 
\label{M_u_t}
\end{equation}
These equations represent the total-mole (mass) conservation equation in the standard form (Eq.\ (\ref{M_t})) and the transport form (Eq.\ (\ref{M_u_t})).
We use one of these equations to replace one of the component conservation equations. 
Here, we replace the water mole (mass) conservation equation with the total-mole (mass) conservation equation.
Even though the rock-fluids system is compressible, the total-mole conservation equation has near-elliptic behavior.
The component conservation equations, on the other hand, have near-hyperbolic behavior. 
% HAT-old: in the absence of Pc, they are hyperbolic. Do you want to say that?

In the following matrix notation, "$P$" refers to the gas-phase pressure, "$R$" refers to the total mole density variable, "$T$" to the total-mole conservation equation, and "$C$" refers
to both the $n_h$ primary component variables and the hydrocarbon conservation equations.
$R_C$ and $R_T$ are then the residuals of the hydrocarbon conservation equations and the total-mole conservation equation, respectively.
$A_{CC}$, $A_{CP}$, $A_{TC}$ and $A_{TP}$ are then 
the derivatives of the hydrocarbon conservation equations versus the primary component variables,
the derivatives of the hydrocarbon conservation equations versus the gas-phase pressure,
the derivatives of the total-mole conservation equation versus the primary component variables and
the derivatives of the total-mole conservation equation versus the gas-phase pressure,
respectively.

\subsection{Sequential Fully Implicit Scheme for Compositional Flow}

We use a finite-volume method with single-point upstream weighting for the spatial discretization and a first-order implicit (backward Euler) scheme for the integration in time.

\subsubsection{Fully implicit method}
The Fully Implicit (FI) method with the natural-variables formulation \cite{Coats:1980} uses the usual form of the conservation equations (\ref{M_c}) and (\ref{M_t})
without the thermodynamic volume equation; that is, the variable $\rho_t$ is not used.
Upon convergence, if $\rho_t$ is computed according to equation (\ref{VOLUME_CONSTRAINT}) , then
we have $S_p^{thermo} = S_p$ for each phase $p\in\{g,o,w\}$.
Algorithm~\ref{FI_method} describes the FI method, in which $\epsilon$ denotes the convergence tolerance.

\bigskip

\begin{algorithm}[H]
   compute $R_C$ and $R_T$ from equations (\ref{M_c}) and (\ref{M_t})\\
   \While{$max(|R_C|, |R_T|)>\epsilon$}{
      compute $A_{CC}$, $A_{CP}$, $A_{TC}$, $A_{TP}$ from equations (\ref{M_c}) and (\ref{M_t})\\
      compute $\left( \begin{array}{c}
\delta C \\
\delta P
\end{array} \right)
 = -\left( \begin{array}{cc}
A_{CC} & A_{CP} \\
A_{TC} & A_{TP}
\end{array} \right)^{-1} 
\left( \begin{array}{c}
R_C \\
R_T
\end{array} \right)$\\
      $C = C + \delta C$\\
      $P = P + \delta P$\\
      compute phase-split only to detect phase appearance\\
      recompute $R_C$ and $R_T$ from equations (\ref{M_c}) and (\ref{M_t})
   }
\caption{FI method}
\label{FI_method}
\end{algorithm}
\bigskip
\noindent

\subsubsection{Nonlinear volume-balance equation}

The temporal discretization of the material balance equations (\ref{M_c}) and (\ref{M_w}) can be written for ${\alpha}=c,w$ as:
\begin{equation}
\left[ \phi \left(\sum_p {\rho}_p S_p\right) z_{\alpha} \right]^{n+1} - \left[ \phi \left(\sum_p {\rho}_p S_p\right) z_{\alpha} \right]^{n}  = \Delta t \tilde{f}_{\alpha}^{n+1}
\label{DISCRETIZED_MATERIAL_BALANCE}
\end{equation}
The contributions from the fluxes and the wells for a hydrocarbon component~$c$ and the water component are:
\begin{eqnarray}
\tilde{f}_c^{n+1} &=& - \nabla \cdot \left[{y}_c{\rho}_g u_g +{x}_c{\rho}_o u_o  \right] + {y}_c{\rho}_g q_g +{x}_c{\rho}_o q_o \\ 
\mbox{and } \tilde{f}_w^{n+1} &=& - \nabla \cdot \left[{\rho}_w u_w \right] + {\rho}_w q_w
\end{eqnarray}
The superscripts $n$ and ${n+1}$ refer to the previous and current timesteps. 
The Sequential Fully Implicit (SFI) nonlinear volume-balance equation for immiscible flow~\cite{Jenny:2006} was obtained as follows: each mole (mass) conservation equation was multiplied 
by the inverse of the molar (mass) density of the phase, and then summed together. When the flow is immiscible,
the inverse of the molar density of the phase is the molar volume of that phase.
The equivalent of the molar volume of each phase for compositional flow is the partial molar volume of each component. 
We define the partial molar volume of component $\alpha$ as:
\begin{equation}
V_{T_{\alpha}}=\frac{\partial V_T}{\partial N_{\alpha}}
\label{VT_ALPHA}
\end{equation}
where $V_T$ is the total fluid volume and $N_{\alpha}$ is the number of moles of component ${\alpha}=c,w$.
In the case of immiscible multiphase flow, where each component can exist in only one phase, the following relationships are satisfied:
\begin{eqnarray}
\sum_c y_c V_{T_c} = \frac{1}{\rho_g} \label{VTC_DO1},\\
\sum_c x_c V_{T_c} = \frac{1}{\rho_o} \label{VTC_DO2}\\
\mbox{and } V_{T_w} = \frac{1}{\rho_w}\label{VTC_DO3}.
\end{eqnarray}
As stated by Watts \cite{Watts:1986}, the partial molar volumes, the overall mole-fractions and the overall specific volume $\frac{1}{\sum_p {\rho}_p S_p}$ satisfy the following relationship:
\begin{eqnarray}
\sum_{{\alpha}=c,w} z_{\alpha} V_{T_{\alpha}} = \frac{1}{\sum_p {\rho}_p S_p}
\label{VTC}
\end{eqnarray}

In this paper, the pressure equation is obtained as the weighted sum of the nonlinear conservation equations (\ref{DISCRETIZED_MATERIAL_BALANCE}); 
the weights are the partial molar volumes. The result can be written as
\begin{eqnarray}
\sum_{{\alpha}=c,w}  V_{T_{\alpha}}^{n+1} \left[ \phi \left(\sum_p {\rho}_p S_p\right) z_{\alpha} \right]^{n+1}
&=& \nonumber \\
\sum_{{\alpha}=c,w}  V_{T_{\alpha}}^{n+1} \left[ \phi \left(\sum_p {\rho}_p S_p\right) z_{\alpha} \right]^{n}
&+& \Delta t \sum_{{\alpha}=c,w}  V_{T_{\alpha}}^{n+1} \tilde{f}_{\alpha}^{n+1}
\end{eqnarray}
Using relationship (\ref{VTC}), it follows that 
\begin{equation}
\sum_{{\alpha}=c,w}  V_{T_{\alpha}}^{n+1} \left[ \phi \left(\sum_p {\rho}_p S_p\right) z_{\alpha} \right]^{n+1} =  \phi^{n+1},
\end{equation}
and the new nonlinear pressure equation can be written as:
\begin{equation}
\phi^{n+1}
= \sum_{{\alpha}=c,w}  V_{T_{\alpha}}^{n+1} \left[ \phi \left(\sum_p {\rho}_p S_p\right) z_{\alpha} \right]^{n}
+ \Delta t \sum_{{\alpha}=c,w}  V_{T_{\alpha}}^{n+1} \tilde{f}_{\alpha}^{n+1}
\label{VOLUME_BALANCE}
\end{equation}
Equation (\ref{VOLUME_BALANCE}) is the nonlinear overall volume-balance. This equation describes
the pressure field that equilibrates the system if the flux terms are taken as implicit functions of
the pressure, but the overall mole-fractions are from the previous composition update. 
In our sequential-implicit solution strategy, we use the Newton method to iterate out the pressure nonlinearities in the overall-volume balance; thus, the derivatives of the partial molar volumes 
with respect to pressure are required. During the pressure solution stage, the overall mole-fraction of each component (in the control volume) is kept fixed.
The constraint equations (\ref{SUMY}) to (\ref{FUGACITY}) are used to get the derivatives of 
$y_c$ and $x_c$ with respect to $P$, and equation (\ref{THERMO_SATURATION_NORMALIZED}) is used to get the derivatives of the saturations with respect to $P$.
After each Newton iteration (pressure solution), a phase-split computation is performed for each cell. Then, the saturations and mole-fractions for each phase are reset to their thermodynamic equilibrium values.
Recent research show that the cost of the phase-split calculations can be reduced to a small fraction of the total simulation cost \cite{MichelsenMollerup:2007,Voskov:2009,Voskov:2017}.
Equation (\ref{VOLUME_BALANCE}) is identical to the pressure equation of the SFI method for the dead-oil (immiscible multiphase flow with no mass transfer),
where the phase densities are used for the decoupling. It is also identical to the decoupling used by \cite{Lee:2008,MoynerBO:2016} for the black-oil formulation 
(i.e., gas component can dissolve in the oil phase).

\subsubsection{Compositional system}
The second step of the sequential implicit method consists of freezing the pressure and total-velocity fields and advecting the components.
For this purpose, we use the transport form of the conservation equations (i.e., Eqs. (\ref{M_u_c}) and (\ref{M_u_t})).

To deal with the transport, two classes of methods have been proposed. The first class uses variables that represent the full fluid content.
This is the case for Acs et al. \cite{Acs:1985}, Watts \cite{Watts:1986}, Trangenstein \& Bell \cite{TBBO:1989,TBC:1989}, 
Dicks \cite{Dicks:1993}, Pau et al. \cite{Pau:2012}, Faigle et al. \cite{Faigle:2014} and Doster et al. \cite{Doster:2014}.
The variables used are: the total number of moles (mass) of each component, or the total number of mole (mass) of each component divided by the pore volume.
This is equivalent to updating the variables $\xi_c$ and $\xi_w$ defined as 
$\xi_c= \phi \frac{\left( {y}_c{\rho}_g S_g +{x}_c{\rho}_o S_o \right)}{\sum_p {\rho}_p S_p} \rho_t$
and 
$\xi_w= \phi \frac{{\rho}_w S_w}{\sum_p {\rho}_p S_p} \rho_t$
and recomputing all the other variables by phase-split computations.

Since these formulations are based on total-mole (total-mass) variables, 
it is not possible to employ a natural-variables (i.e., saturations-based) formulation to skip the phase-split computations.
In the natural-variables formulation, the local equilibrium constraints Eqs.\ (\ref{FUGACITY}) are solved
simultaneously with the conservation equations (\ref{M_u_c}) and (\ref{M_u_t}) by the Newton method, 
and the phase-split computations are only used to determine if a fluid phase has appeared.
After convergence of the coupled system of component conservation equations, the moles (mass) of each component are conserved; however,  
the thermodynamic volume equation (\ref{VOLUME_CONSTRAINT}) may not be satisfied exactly.
Since this error is local, Acs et al. \cite{Acs:1985}, Watts \cite{Watts:1986}, 
Pau et al. \cite{Pau:2012}, Faigle et al. \cite{Faigle:2014} and Doster et al. \cite{Doster:2014} proposed 
to accept the timestep, but to introduce a local relaxation term in order to keep the error bounded in time.
Dicks \cite{Dicks:1993} studied the thermodynamic volume error and introduced a smoothing algorithm that locally perturbs the material balance.
Trangenstein \& Bell \cite{TBBO:1989,TBC:1989} mentioned that the pressure equation is a linearization of the thermodynamic volume error,
and that it could be possible to add an outer-loop on the pressure and the compositional system.
However, in their applications, they accept the timestep after one outer-loop.
No study exists of the convergence rate of the outer iterations.

The second class of methods \cite{Jenny:2006,Lee:2008,Hajibeygi:2014,MoynerBO:2016,Moncorge:2017,MoynerRSC:2017,MoynerIPAM:2017} uses saturations and mole- or mass-fractions.
Since the composition variables are only fractions, it is necessary to remove one conservation equation during the solution of the coupled transport.
It then means that the pressure equation is not considered a volume-balance, but rather a total-mole or total-mass conservation law.
% rephrase
In immiscible situations, with the exception of small time scale transient situations, flow and transport usually can be considered decoupled.
But for compositional systems, flow and transport usually cannot be decoupled, and it is thus necessary to perform several outer iterations
to reduce to an acceptable tolerance.
Recently, this second class of methods has been abandoned, that is, the first class of methods is preferred \cite{MoncorgeSIAM:2017,MoynerSIAM:2017}.

Here, we propose to converge the system as in the first class of methods with our new extended natural-variables formulation.
%
% HAT disjointed
The pressure and the total-velocity are frozen. The total mole density is used as an independent variable, and we 
converge the transport system with the $n_h+1$ conservation equations (\ref{M_u_c}) and (\ref{M_u_t}).
The variables are the $n_h$ primary component variables and the additional total mole density variable $\rho_t$.
The advantage of this formulation - compared with using the number of moles, or mass, per component - is that
for the compositional update, we retain the benefits of the natural-variables formulation. Namely, we
better control the nonlinearities with the saturations as variables and we compute the phase-split calculations only when a new phase is detected.
At convergence, all the components are conserved to the prescribed tolerance; however, the thermodynamic volume equation may not be satisfied. 
We define the total-velocity error as the difference between the total-velocity based on the latest variables after the compositional update and the total-velocity obtained from the pressure solution and frozen during the compositional update. We show that the total-velocity error, which had not been identified before, is present in the invaded regions, and that it is the primary source of nonlinear convergence problems associated with the splitting scheme. This total-velocity error is particularly important in regions experiencing appearance/disappearance of the gas phase during the timestep.
The thermodynamic volume residual is defined as 
\begin{equation}
R_{thermo}=\sum_p S_p^{thermo} - 1.0
\end{equation}
and 
the relative total-velocity residual as 
\begin{equation}
R_{u_t}=\frac{1}{u_t} ( u_t^f - u_t),
\end{equation}
where $u_t^f$ is the total-velocity, which is kept fixed during the composition update, and $u_t$ is the total-velocity computed by equation (\ref{UT}) with the latest updated variables.
Upon solving the nonlinear pressure equation, two strategies are possible.
The first strategy consists of continuing with outer iterations until the dimensionless residuals $|R_{thermo}|_{\infty}$ and $|R_{u_t}|_{\infty}$ (infinity norm) fall below a tight tolerance. 
For strongly coupled flow and transport, even a tolerance of 0.01 may take many outer iterations, 
% HAT confusing
or no convergence may be achieved at all.
This strategy is not practical, as it requires too many iterations in order to be competitive with methods like mSFI \cite{Moncorge:2017}.
The second strategy consists of relaxing the tolerances of $|R_{thermo}|_{\infty}$ and $|R_{u_t}|_{\infty}$ in order to achieve convergence with less outer iterations.
Another measure of the total-velocity error is based on the divergence of the total-velocity.
The dimensionless divergence of the total-velocity is written as
\begin{equation}
\overline{\nabla \cdot  u_t} = \frac{\sum_j \left(Q_{t}\right)^{j,i}}{\max(\frac{(V_P)^i}{\Delta t}, \, \max_j(|\left(Q_{t}\right)^{j,i}|))}
\end{equation}
where~$j$ are the cells surrounding cell~$i$. 
$\left(Q_{t}\right)^{j,i}$ are the total volumetric rate from cell~$j$ to cell~$i$ (positive if entering cell~$i$, negative if leaving cell~$i$), 
$(V_P)^i$ the pore volume of cell i and $\Delta t$ the discretized timestep.
The divergence of the total-velocity error is defined by the dimensionless residual
\begin{equation}
R_{\nabla \cdot u_t}=\frac{\sum_j \left(Q^f_{t}\right)^{j,i} - \sum_j \left(Q_{t}\right)^{j,i}}{\max(\frac{(V_P)^i}{\Delta t}, \, \max_j(|\left(Q^f_{t}\right)^{j,i}|), \, \max_j(|\left(Q_{t}\right)^{j,i}|))}
\end{equation}
where $\left(Q^f_{t}\right)^{j,i}$ are the total volumetric rates from cell~$j$ to cell~$i$ that have been frozen during the composition update.

For challenging test cases, we observe that tolerances of 
0.2 for $|R_{thermo}|_{\infty}$, 
0.4 for $|R_{u_t}|_{\infty}$ and 
0.1 for $|R_{\nabla \cdot u_t}|_{\infty}$
are good enough for strongly coupled flow-transport problems.
From this point on, we can accept the timestep with the converged accumulation terms computed with the quantities $\xi_c$ and $\xi_w$.
The accepted volume balance error: $\sum_p \left( S_p^{thermo} - S_p \right)$ 
% HAT rephrase
can be added to the pressure system in order to not propagate this error.
%
% HAT: vague
However, the volume-balance equation (\ref{VOLUME_BALANCE}) already prevents the propagation of the volume-balance error that potentially exists at time $t^n$ by equilibrating volumes at time $t^{n+1}$.
Adding the volume-balance error does not change the results, but may perturb the solution in some situations. We do not recommend its use.

The algorithm of this second strategy is described in Algorithm~\ref{SFI_method}.

\begin{algorithm}[H]
   \bigskip
\While{$\max(|R_C|_{\infty}, |R_T|_{\infty})>\epsilon$ or $|R_{thermo}|_{\infty}>\epsilon_{thermo}$ or $|R_{u_t}|_{\infty}>\epsilon_{u_t}$ or $|R_{\nabla \cdot u_t}|_{\infty}>\epsilon_{\nabla \cdot u_t}$}{
  freeze $z_{\alpha=c,w}$\\
  \While{$|\delta P|_{\infty}>\epsilon_P$}{
      compute derivative wrt P and residual of volume-balance equation (\ref{VOLUME_BALANCE}) with derivatives of $y_c$ and $x_c$ given by constraints (\ref{SUMY}) to (\ref{FUGACITY}) and derivatives of $S_p$ by (\ref{THERMO_SATURATION_NORMALIZED}) \\
      solve and update $P = P + \delta P$\\
      compute phase-split and set $y_c$, $x_c$ and $S_p$ to thermodynamic values with (\ref{PHASE_SPLIT_SYSTEM_ZC}) to (\ref{THERMO_SATURATION_NORMALIZED})\\
      recompute residual of volume-balance equation (\ref{VOLUME_BALANCE})\\
   }
   unfreeze $z_{\alpha=c,w}$\\
   compute and freeze total-velocity $u_t$ \\
   \While{$\max(|\delta C|_{\infty}, |\delta \rho_t|_{\infty})>\epsilon_C$}{
      compute Jacobian and residual of equations (\ref{M_u_c}) and (\ref{M_u_t}) (all conservation equations) wrt variables $C$ and $\rho_t$; constraints equations (\ref{SUMY}) to (\ref{FUGACITY}) are used to convert the derivatives wrt secondary variables in function of derivatives wrt primary variables $C$\\
      solve and update $C = C + \delta C$ and $\rho_t = \rho_t + \delta \rho_t$\\
      compute phase-split only to detect phase appearance\\
     compute $R_C$ and $R_T$ from equations (\ref{M_u_c}) and (\ref{M_u_t})\\
   }
   compute $R_{thermo}$, $R_{u_t}$ and $R_{\nabla \cdot u_t}$\\
   unfreeze total-velocity $u_t$ \\
}
Exact conservative quantities are $\xi_c$ and $\xi_w$\\
   \bigskip
\caption{Compositional SFI method}
\label{SFI_method}
\end{algorithm}

\vspace{20pt}

In the case of immiscible multiphase flow with one component per phase, our solution and the solution of Jenny et al. \cite{Jenny:2006} are identical.
% HAT vague
This is not the case for black-oil formulations, where the solution of Lee et al. \cite{Lee:2008} and M{\o}yner \& Lie \cite{MoynerBO:2016} is relaxing the material balance of one component.

\subsubsection{Comparison with other compositional pressure equations}

Similar to the nonlinear volume-balance equation, we use
the gas-phase pressure $P$ and the overall mole-fractions $z_{\alpha}$ for each component $\alpha=c,w$.
The overall mole-fractions $z_{\alpha}$ are fixed, and the constraint equations (\ref{SUMY}) to (\ref{FUGACITY}) are used to get the derivatives of $y_c$ and $x_c$ wrt $P$. Equation (\ref{THERMO_SATURATION_NORMALIZED}) is used to get the derivatives of the saturations with respect to $P$.
Sequential formulations for compositional flows where derived in a semi-implicit context by 
Acs et al. \cite{Acs:1985}, Watts \cite{Watts:1986}, Trangenstein \& Bell \cite{TBBO:1989,TBC:1989}, Dicks \cite{Dicks:1993},
Pau et al. \cite{Pau:2012}, Faigle et al. \cite{Faigle:2014} and Doster et al. \cite{Doster:2014}.
In these previous methods, the pressure equation is derived by linearizing the difference between the sum of the thermodynamic volumes of each phase
and the pore volume. The resulting pressure equation accounts for the change of volume of the fluids due to the advection of the compositions.
We refer to \cite{Watts:1986} for the derivation of the `linearized volume-balance pressure equation' that can be written 
in semi-discretized in time form 
\begin{equation}
\left[ \frac{d \phi}{d P} + \frac{\phi}{\sum_p \rho_p S_p} \frac{\partial \sum_p \rho_p S_p}{\partial P} \right] (P^{n+1} - P^{n})
=
\Delta t \sum_{{\alpha}=c,w}  V_{T_{\alpha}} \tilde{f}_{\alpha}^{n+1}.
\end{equation}
In a semi-implicit situation, we would first converge
\begin{equation}
\left[ \frac{d \phi}{d P} + \frac{\phi}{\sum_p \rho_p S_p} \frac{\partial \sum_p \rho_p S_p}{\partial P} \right]^{n} (P^{n+1} - P^{n})
=
\Delta t \sum_{{\alpha}=c,w}  V_{T_{\alpha}}^{n} \tilde{f}_{\alpha}^{n+1}
\end{equation}
and then switch to the nonlinear composition update. 
In a sequential implicit situation we would converge the equation
\begin{equation}
\left[ \frac{d \phi}{d P} + \frac{\phi}{\sum_p \rho_p S_p} \frac{\partial \sum_p \rho_p S_p}{\partial P} \right]^{\nu} (P^{\nu+1} - P^{n})
=
\Delta t \sum_{{\alpha}=c,w}  V_{T_{\alpha}}^{\nu} \tilde{f}_{\alpha}^{\nu+1}
\label{LINEARIZED_PRESSURE_EQUATION}
\end{equation}
until $|P^{\nu+1}-P^{\nu}|$ is small, where the superscripts $\nu$ and  $\nu+1$ denote the old and new pressure iteration levels.
Coats \cite{Coats:2000} built a pressure equation by algebraic manipulations
of the usual material balance equations, and M{\o}yner et al. \cite{MoynerIPAM:2017,MoynerSIAM:2017} 
use this pressure for their sequential implicit method for compositional flow.
Coats' pressure equation in a sequential implicit setting is equivalent to converging the equation
\begin{eqnarray}
\sum_{{\alpha}=c,w}  V_{T_{\alpha}}^{\nu} \left[ \phi \left( \sum_p \rho_p S_p \right) z_{\alpha} \right]^{\nu+1}
&=& \nonumber \\
\sum_{{\alpha}=c,w}  V_{T_{\alpha}}^{\nu} \left[ \phi \left( \sum_p \rho_p S_p \right) z_{\alpha} \right]^{n}
&+&
\Delta t \sum_{{\alpha}=c,w}  V_{T_{\alpha}}^{\nu} \tilde{f}_{\alpha}^{\nu+1}
\label{COATS_PRESSURE_EQUATION}
\end{eqnarray}
until $|P^{\nu+1}-P^{\nu}|$ is small, where the superscripts $\nu$ and  $\nu+1$ denote the old and new pressure iteration levels, and $z_{\alpha}$ in the left-hand-side term is taken at the latest composition iteration level.

The overall compressibility of the system can dramatically change during the transition from a state with only liquid phases present to a state where a gas phase is present. 
Moreover, $V_{T_{\alpha}}$ can be very strong functions of pressure as seen in the relationships (\ref{VTC_DO1}), (\ref{VTC_DO2}) and (\ref{VTC_DO3}).
It is then possible to encounter cases where the pressure equations (\ref{LINEARIZED_PRESSURE_EQUATION}) and (\ref{COATS_PRESSURE_EQUATION}) do not converge. 
It is then necessary to use both the derivatives of the compressibility and the partial molar volumes.
The linearized equation (\ref{LINEARIZED_PRESSURE_EQUATION}) and the nonlinear volume-balance equation (\ref{VOLUME_BALANCE}) are only equivalent for incompressible immiscible systems..
The Coats pressure equation (\ref{COATS_PRESSURE_EQUATION}) and the nonlinear volume-balance equation (\ref{VOLUME_BALANCE}) are different because $V_{T_{\alpha}}$
depend on pressure, and the left-hand-side of (\ref{COATS_PRESSURE_EQUATION}) is only equals to $\phi^{n+1}$ at convergence.
However, when the Coats' pressure equation converges, the solution is the same as the nonlinear volume-balance equation.

\section{Results and Discussion}

The nonlinear strategies used for the numerical experiments with the FI method, the volume balance system, and the compositional SFI method
ensure that the variations of pressure, saturations and mole-fractions over a Newton iteration get 
damped to respectively 1000.0 Psi, 0.2 and 0.2, and 
that the saturations and mole-fractions are kept between 0.0 and 1.0.
If the value of the total mole density $\rho_t$ after a Newton iteration is negative, it is set to zero,
and if the variation of $\rho_t$ over a Newton iteration is larger than the maximum allowed value, i.e., if $|\delta \rho_t|_{\infty} > \max_p \rho_p$, 
the value of $\rho_t$ is set to $\rho_t = \sum_p \rho_p S_p$.

\subsection{1D test cases}
In this section we test our compositional SFI algorithm using 1D models; first for dead-oil and then for live-oil fluids.

\subsubsection{Dead-oil}
\begin{figure}
    \centering
    \begin{subfigure}[h!]{0.8\textwidth}
        %\centering
        \includegraphics[width=\textwidth]{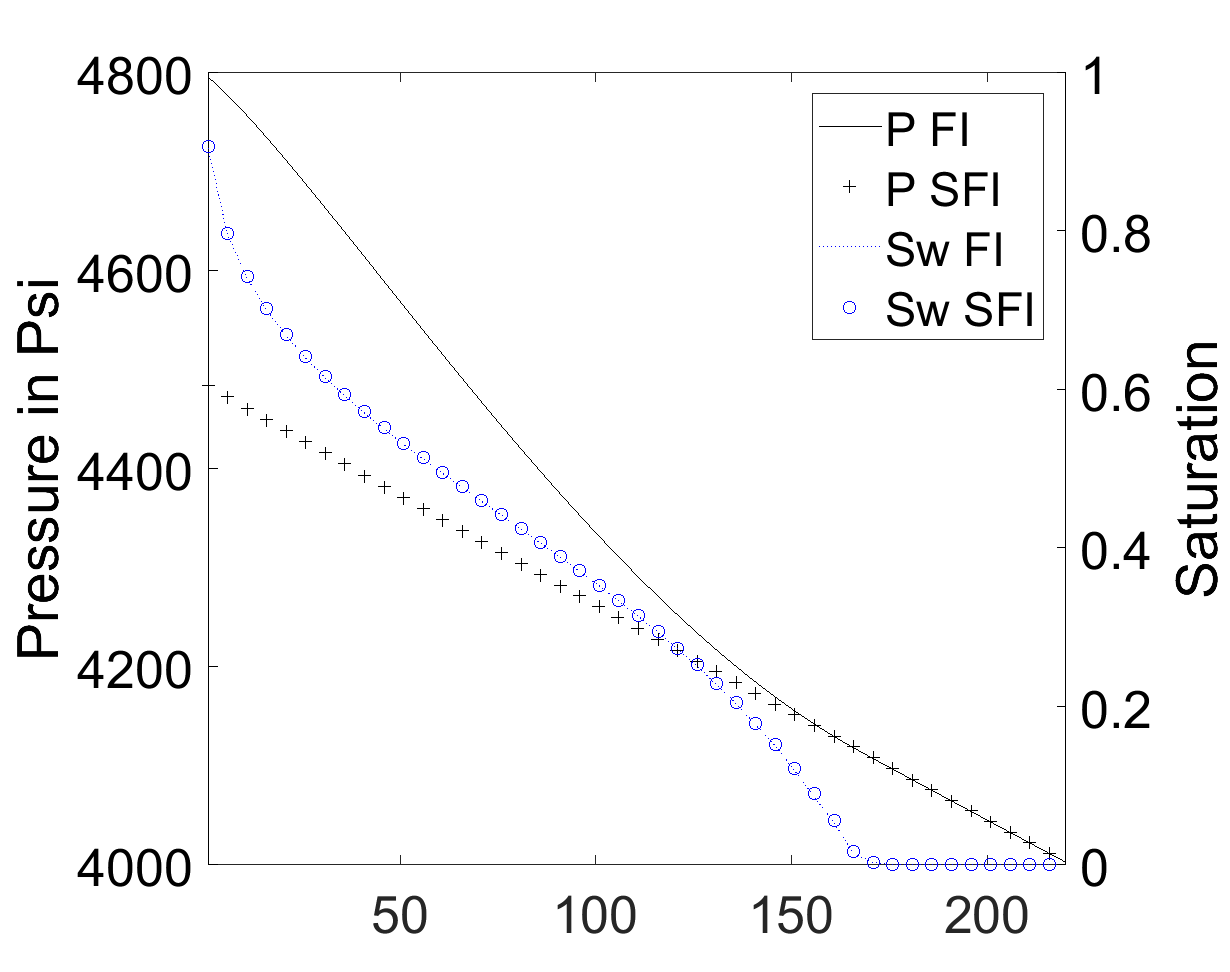}
        \caption{SFI profiles after one outer iteration and converged FI profiles.}
        \label{INJW_1D_LARGEDT_TRANSBELL1}
    \end{subfigure}
    \\
    \begin{subfigure}[h!]{0.8\textwidth}
        %\centering
        \includegraphics[width=\textwidth]{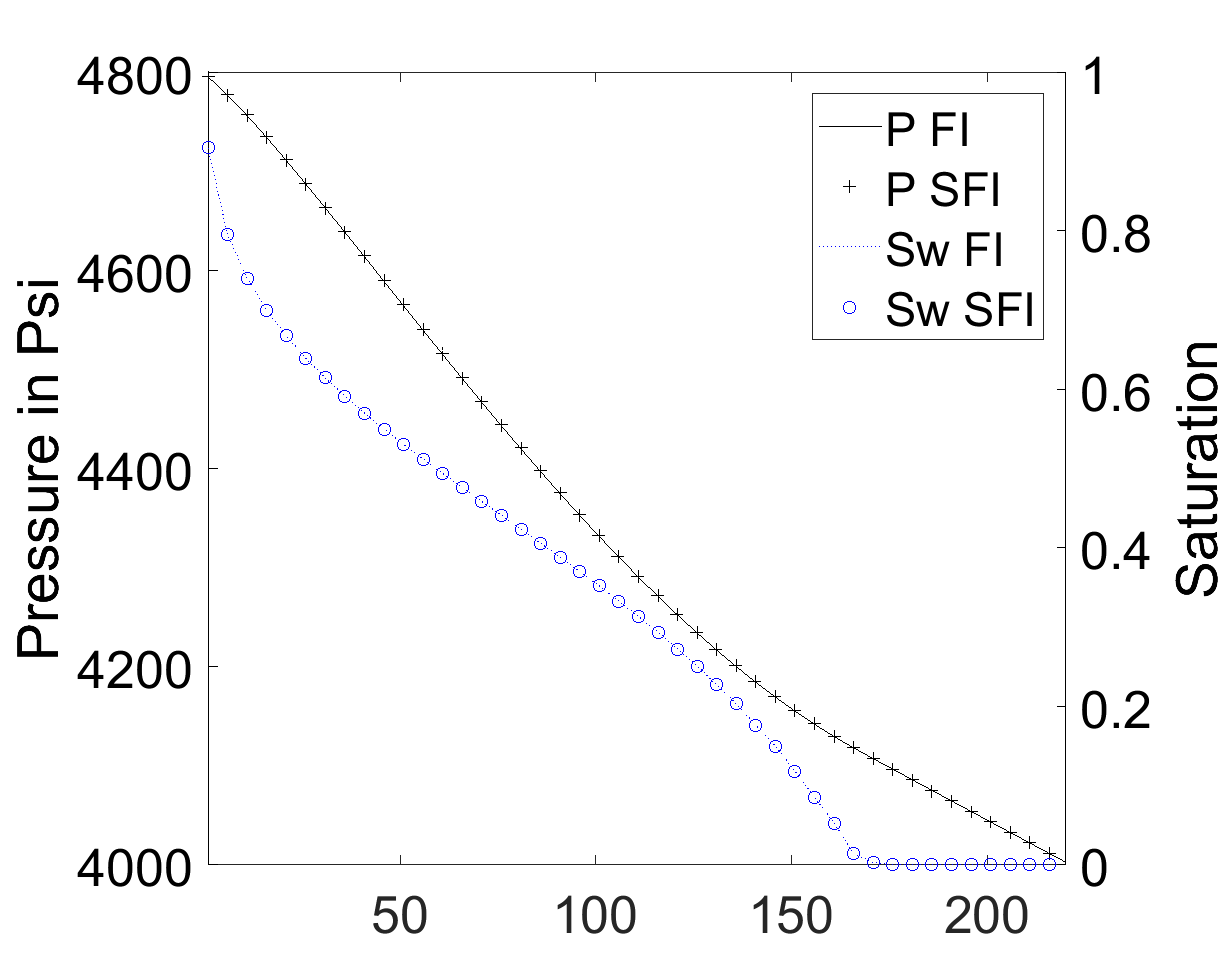}
        \caption{Converged SFI and FI profiles.}
        \label{INJW_1D_LARGEDT_TRANSBELL}
    \end{subfigure}
    \caption{Water injection: pressure and water saturation profiles versus cell number for the FI and the SFI methods.}
    \label{INJW_1D_LARGEDT}
\end{figure}

\begin{figure}
    \centering
    \includegraphics[width=0.7\textwidth]{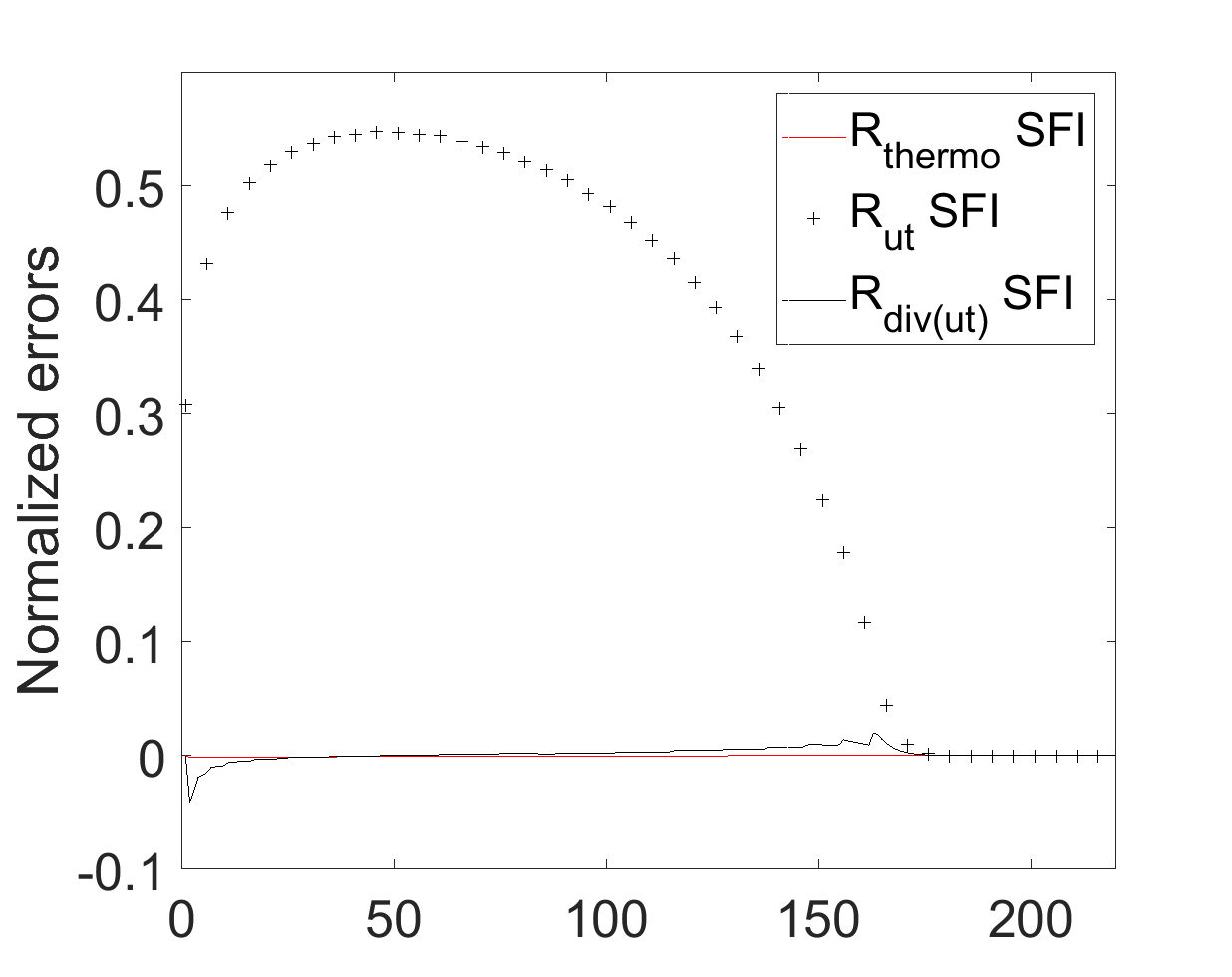}
    \caption{Water injection: thermodynamic volume error $R_{thermo}$, total-velocity error $R_{u_t}$ and divergence of the total-velocity error $R_{\nabla \cdot u_t}$ for the SFI method after one outer iteration.}
    \label{INJW_1D_LARGEDT_TRANSBELL1_ERROR}
\end{figure}

First, the compositional SFI algorithm is tested for cases with no mass transfer between the fluid phases.
We consider injection of water into a reservoir saturated with dead-oil.
With a CFL number of 140 \cite{Coats:2003CFL}, the FI method converges in 16 Newton iterations for the first timestep.
The SFI method was stopped after one outer iteration of Algorithm~\ref{SFI_method}.
The first outer iteration of the SFI method consists of three pressure iterations and 16 compositional iterations.
The results are shown is Figure \ref{INJW_1D_LARGEDT_TRANSBELL1}.
The saturation fronts obtained with the FI and SFI methods are in very good agreement, but the pressure profiles are quite different.
For the SFI method, the material balance equations are satisfied, but the splitting errors 
$|R_{thermo}|_{\infty}$, $|R_{u_t}|_{\infty}$ and $|R_{\nabla \cdot u_t}|_{\infty}$
are, respectively, 2e-3, 0.55 and 0.04.
Figure \ref{INJW_1D_LARGEDT_TRANSBELL1_ERROR} shows the profiles of the three splitting errors.
$R_{thermo}$ is very small, $R_{u_t}$ has large values across the 1D domain, and
$R_{\nabla \cdot u_t}$ is around 1\% except for some localized areas.
The total-velocity errors have an impact on the absolute value of the pressure; however, since the system is immiscible and
% HAT rephrase 
the divergence of the total-velocity is correctly reconstructed,
the transport is already correct during the first outer iteration.
Two outer SFI iterations (a total of 6 pressure iterations and 18 compositional iterations) are required to fully converge the system with 
$|R_{thermo}|_{\infty}$, $|R_{u_t}|_{\infty}$ and $|R_{\nabla \cdot u_t}|_{\infty}$
set to 1e-5, 1e-2 and 6e-3, respectively. These results are shown in Figure \ref{INJW_1D_LARGEDT_TRANSBELL}, and one can observe that the profiles 
obtained with the FI and SFI methods are identical.

\begin{figure}
    \centering
    \begin{subfigure}[h!]{0.8\textwidth}
        %\centering
        \includegraphics[width=\textwidth]{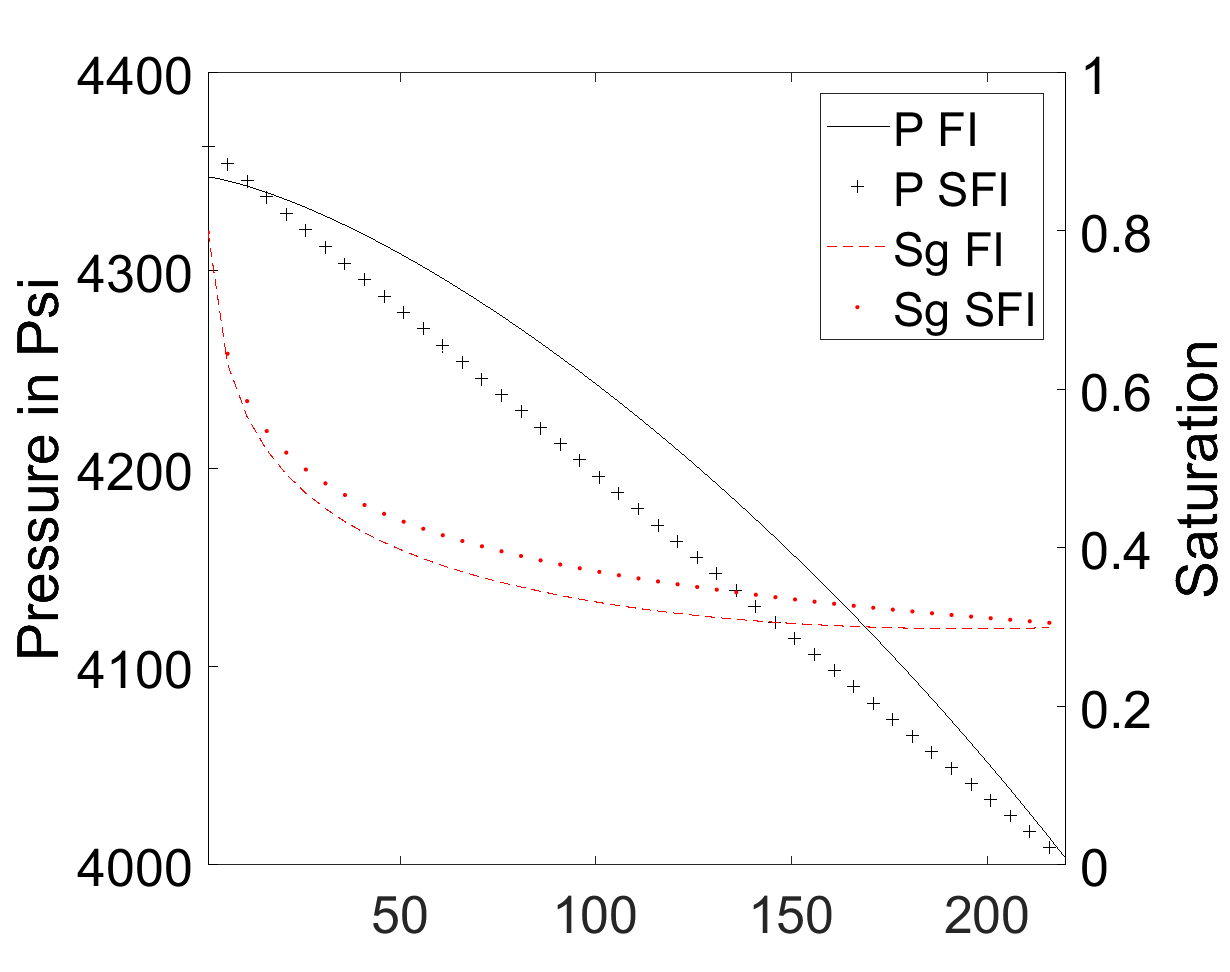}
        \caption{SFI profiles after one outer iteration and converged FI profiles.}
        \label{INJDG_1D_LARGEDT_TRANSBELL1}
    \end{subfigure}
    \\
    \begin{subfigure}[h!]{0.8\textwidth}
        %\centering
        \includegraphics[width=\textwidth]{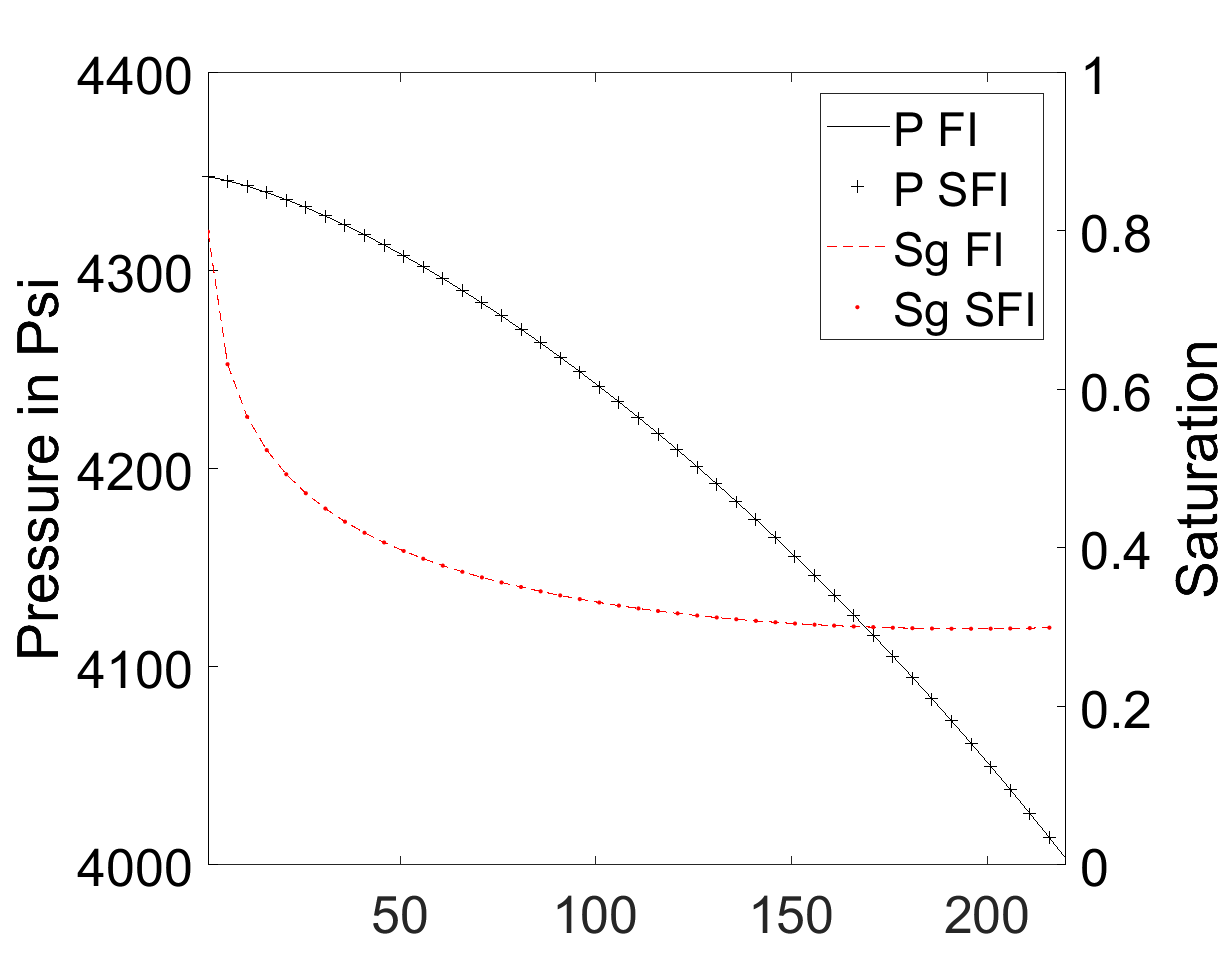}
        \caption{Converged SFI and FI profiles.}
        \label{INJDG_1D_LARGEDT_TRANSBELL}
    \end{subfigure}
    \caption{Injection of dry-gas: pressure and gas saturation profiles versus cell number for the FI and the SFI methods.}
    \label{INJDG_1D_LARGEDT}
\end{figure}

In the second test case, injection of dry-gas into a dead-oil reservoir is considered.
The gas, which is less dense,  less viscous, and more compressible than the oil phase,
is injected with a CFL number of 780.
The ratio of the oil to gas mass densities is 735; the ratio of the oil to gas viscosities is 8.1, and the ratio of the gas to oil compressibilities is 0.005.
% Densities are 35.27 and 0.048 lb/ft3
% Viscosities are 0.243 and 0.03 cP 
% Compressibilities are 1.4e-5 and 3e-3 1/Psi 
For the first timestep, the FI method converges in 42 Newton iterations.
The SFI method was stopped after one outer iteration.
The results are shown in Figure \ref{INJDG_1D_LARGEDT_TRANSBELL1}. The figure shows that the 
pressure and saturation profiles obtained with the FI and SFI methods are quite different.
The first outer iteration of the SFI method consists of eight pressure iterations and seven compositional iterations
($|R_{thermo}|_{\infty}$, $|R_{u_t}|_{\infty}$ and $|R_{\nabla \cdot u_t}|_{\infty}$ are 0.40, 0.81 and 0.12).
Three outer SFI iterations consisting of 18 pressure and 12 compositional iterations are required to converge the SFI method.
The results are shown in Figure \ref{INJDG_1D_LARGEDT_TRANSBELL}, and it can be observed that the profiles obtained with the FI and SFI methods are identical.

\begin{figure}
    \centering
    \begin{subfigure}[h!]{0.8\textwidth}
        %\centering
        \includegraphics[width=\textwidth]{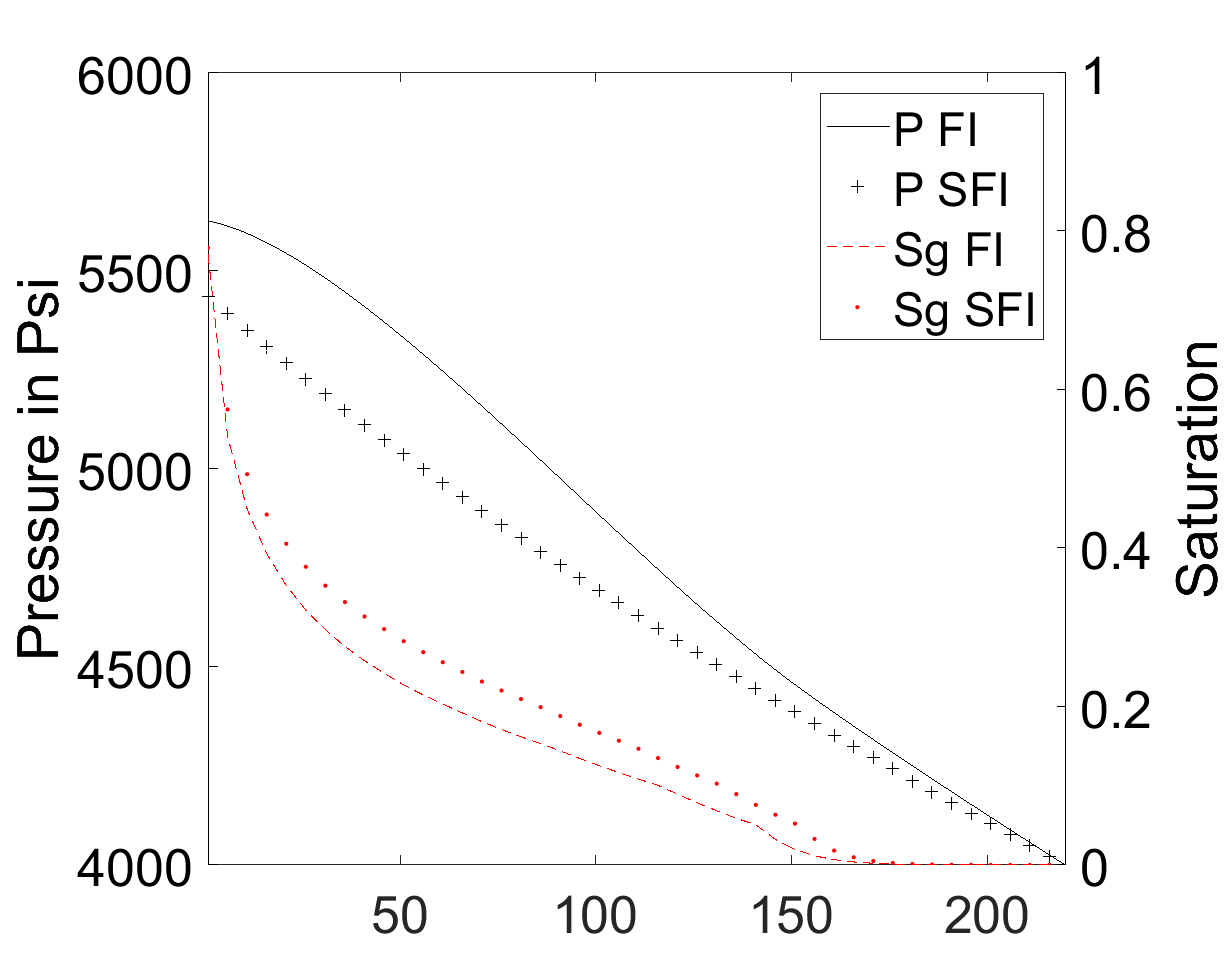}
        \caption{SFI profiles after one outer iteration and converged FI profiles.}
        \label{INJDG_1D_GRAV_LARGEDT_TRANSBELL1}
    \end{subfigure}
    \\
    \begin{subfigure}[h!]{0.8\textwidth}
        %\centering
        \includegraphics[width=\textwidth]{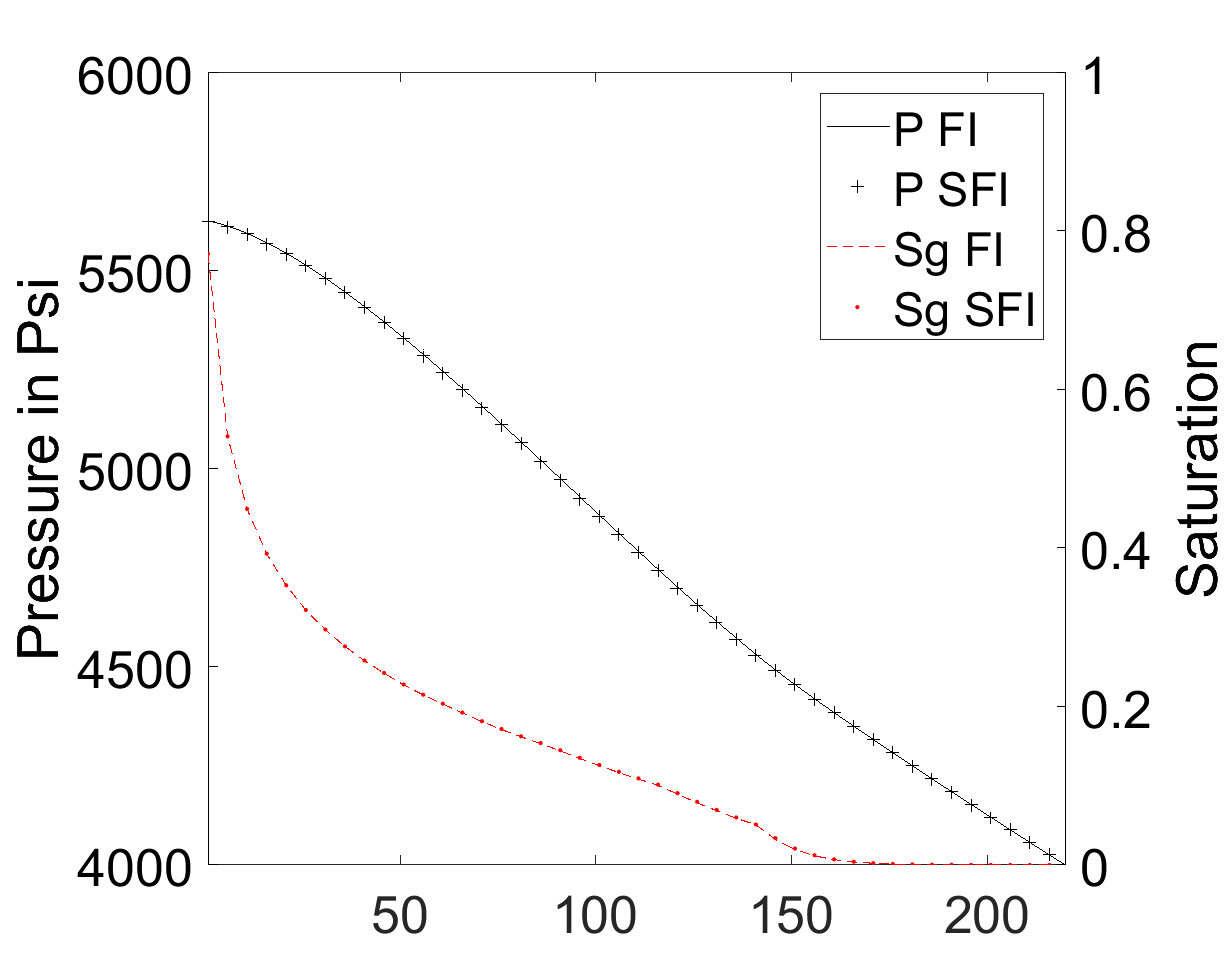}
        \caption{Converged SFI and FI profiles.}
        \label{INJDG_1D_GRAV_LARGEDT_TRANSBELL}
    \end{subfigure}
    \caption{Injection of dry-gas with gravity: pressure and gas saturation profiles versus cell number for the FI and the SFI methods.}
    \label{INJDG_1D_GRAV_LARGEDT}
\end{figure}

As a last dead-oil case, gas injection with gravity is considered.
Gas is injected into a reservoir initially filled with pure oil, 
and a CFL number of 65 was applied.
The FI method converged in 16 Newton iterations for the first timestep.
The SFI method was stopped after one outer iteration.
The results are shown in Figure \ref{INJDG_1D_GRAV_LARGEDT_TRANSBELL1}, which depicts pressure and saturation
profiles as functions of depth.
The pressure and saturation profiles obtained with the FI and the SFI methods are quite different.
The first outer iteration of the SFI method consists of 14 pressure iterations and eight compositional iterations
($|R_{thermo}|_{\infty}$, $|R_{u_t}|_{\infty}$ and $|R_{\nabla \cdot u_t}|_{\infty}$ are 0.25, 0.73 and 0.16).
Three outer SFI iterations consisting of 27 pressure and 13 compositional iterations are required to converge the SFI method.
For this test case with initially pure oil in the reservoir, it is not possible to reach convergence
if the usual linearized pressure equation (\ref{LINEARIZED_PRESSURE_EQUATION}) or the Coats pressure equation (\ref{COATS_PRESSURE_EQUATION}) are employed.
The results are shown in Figure \ref{INJDG_1D_GRAV_LARGEDT_TRANSBELL}, and it can be observed that the profiles obtained with the FI and SFI methods are identical.

\subsubsection{Depletion of a live-oil reservoir}
\begin{figure}
    \centering
    \includegraphics[width=0.7\textwidth]{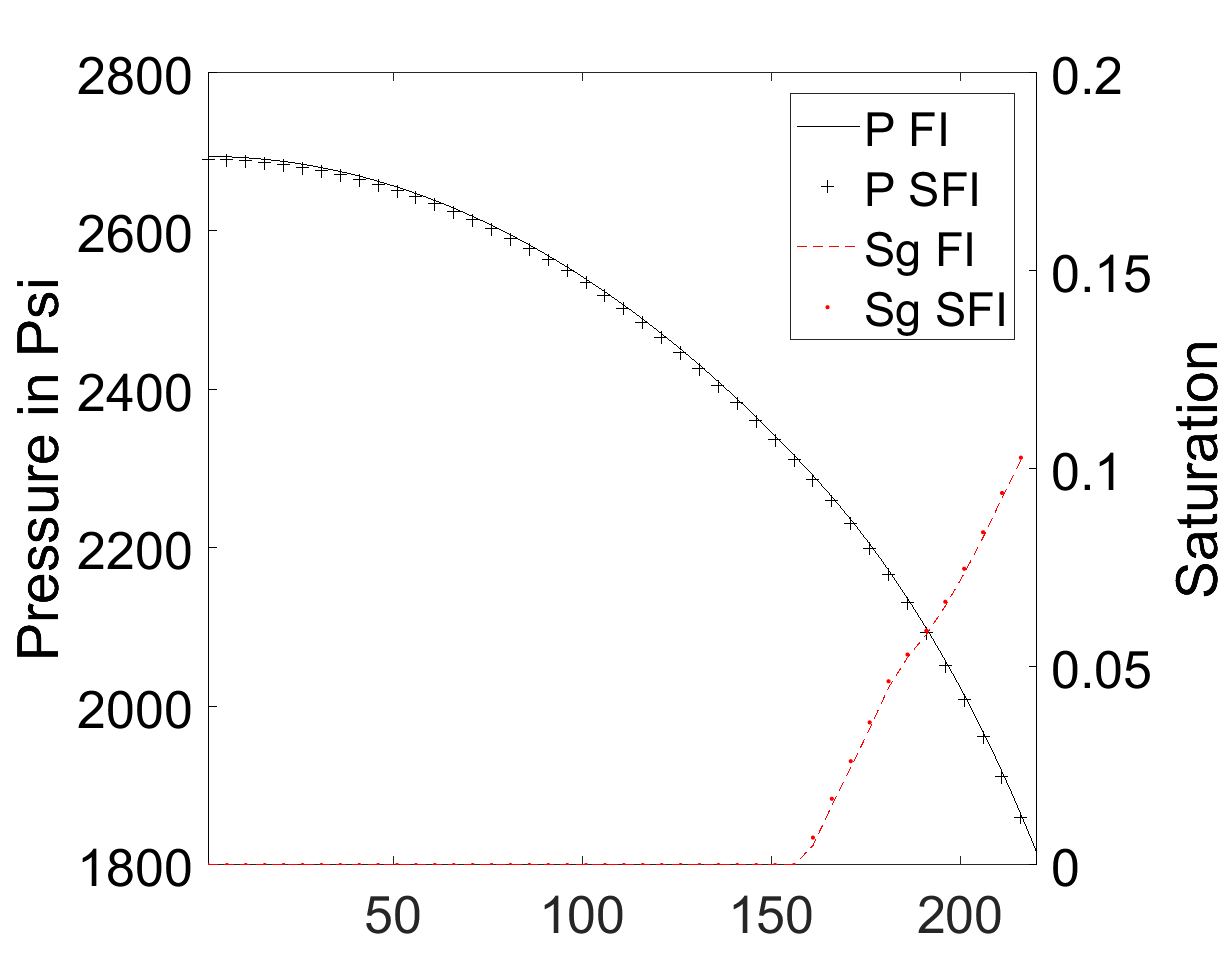}
    \caption{Depletion of live oil: pressure and gas saturation profiles versus cell number for the FI and the SFI methods.}
    \label{DEPL_1D}
\end{figure}

Now that the algorithm is validated for dead-oil cases, we consider a depletion case with the SPE~5 fluid \cite{SPE5}.
The reservoir is originally saturated with pure oil, and we drop the pressure at the right-side. 
In one timestep with a CFL number of 56, gas appears on the right side of the reservoir.
The FI method requires seven iterations to converge.
One outer iteration of the SFI method consisting of six pressure and three compositional iterations was employed
($|R_{thermo}|_{\infty}$, $|R_{u_t}|_{\infty}$ and $|R_{\nabla \cdot u_t}|_{\infty}$ are 0.36, 0.07 and 0.05).
The pressure and gas saturation profiles for the FI and SFI methods are shown in Figure \ref{DEPL_1D}.
The SFI method converges to the correct solution in one outer iteration even though the thermodynamic volume error of 0.36.
However, $R_{thermo}$ is  large for a single cell where the gas phases appears. 
%This localized error does not impact the solution
In two outer iterations (consisting of nine pressure and four compositional iterations), $|R_{thermo}|_{\infty}$, $|R_{u_t}|_{\infty}$
and $|R_{\nabla \cdot u_t}|_{\infty}$ decrease to, respectively, 1e-3, 2e-3 and 2e-3.

\subsubsection{Gas injection in a live-oil reservoir}

\begin{figure}
    \centering
    \begin{subfigure}[h!]{0.8\textwidth}
        %\centering
        \includegraphics[width=\textwidth]{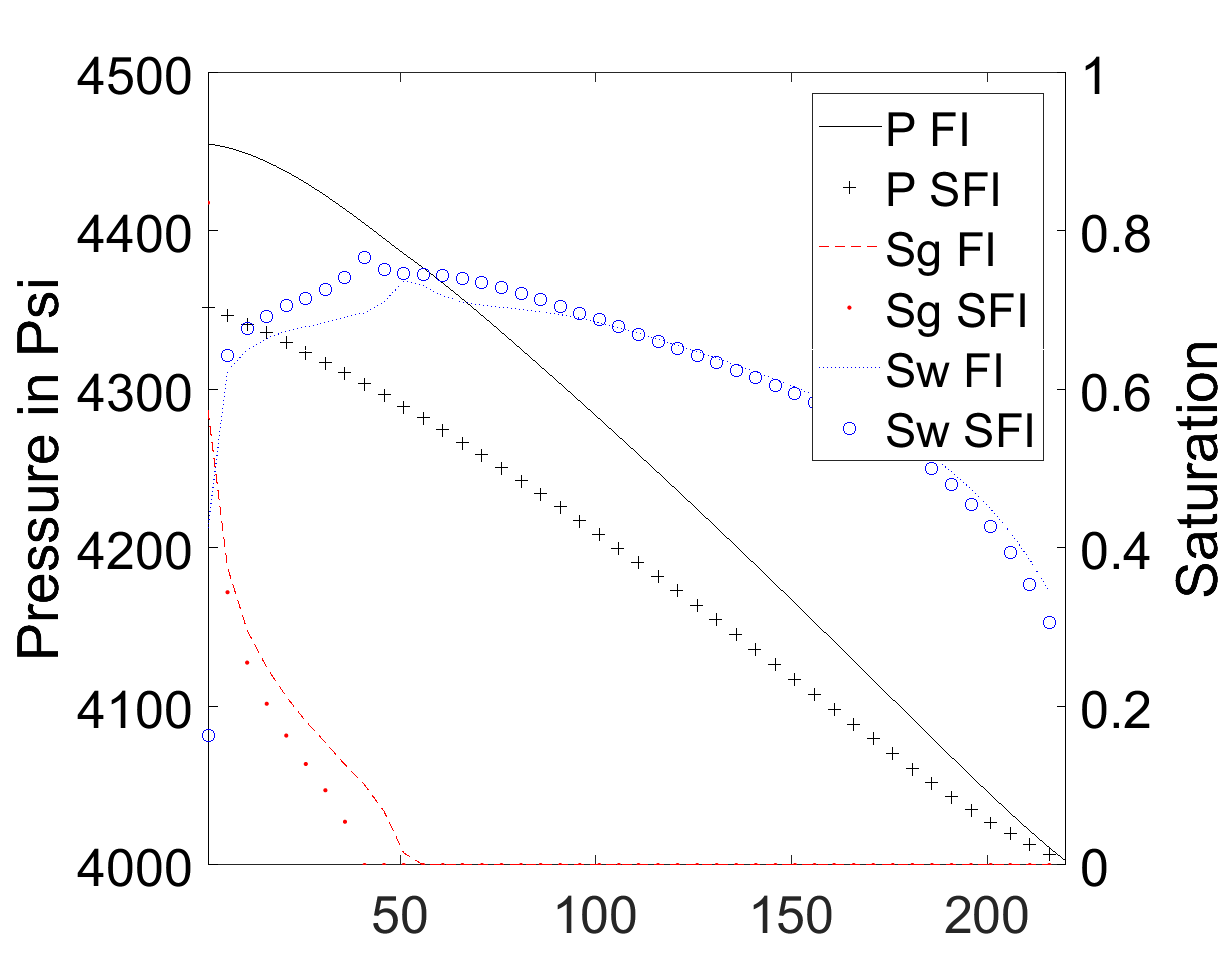}
        \caption{SFI profiles after one outer iteration and converged FI profiles.}
        \label{INJG_1D_GOW_TRANSBELL1}
    \end{subfigure}
    \\
    \begin{subfigure}[h!]{0.8\textwidth}
        %\centering
        \includegraphics[width=\textwidth]{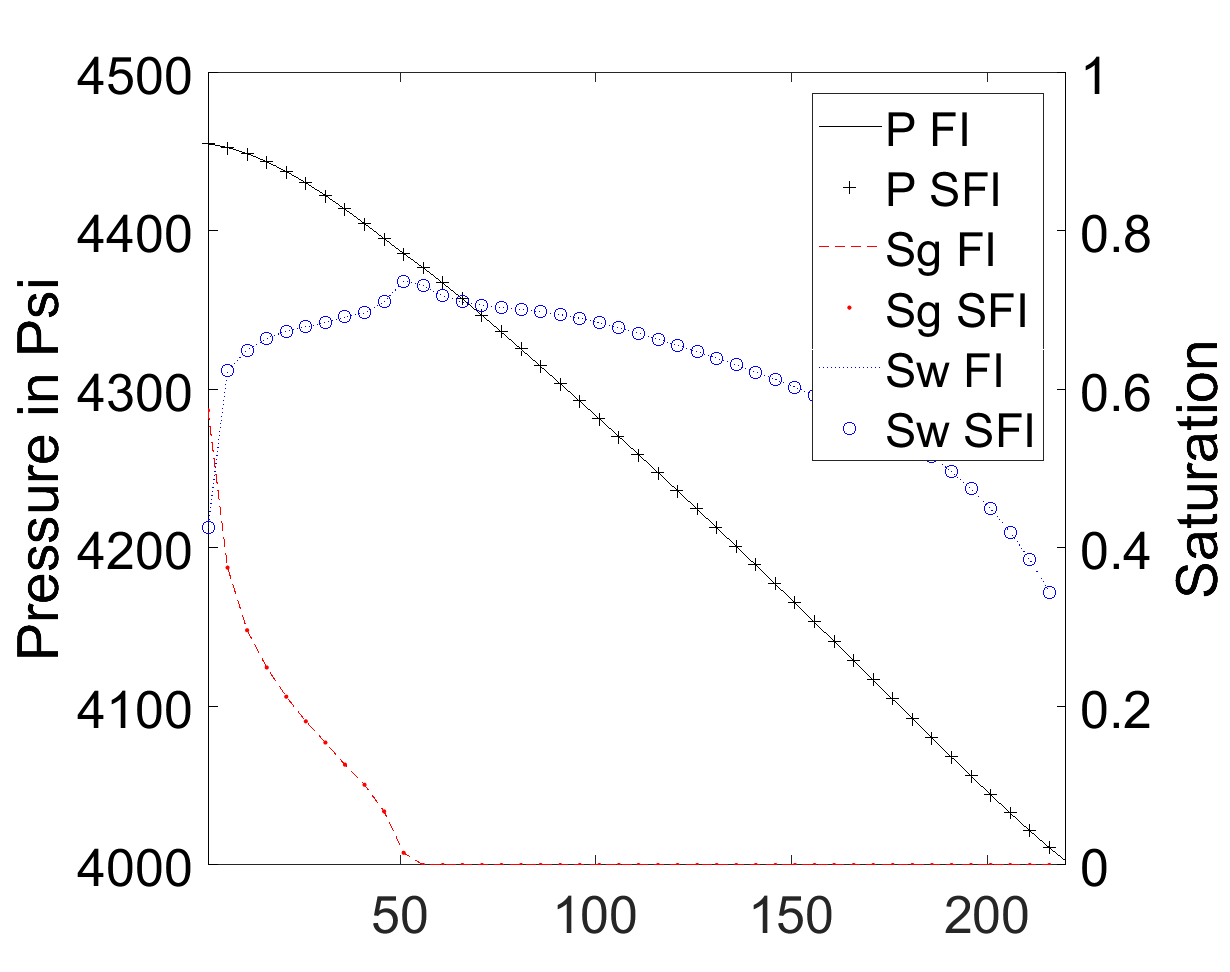}
        \caption{Converged SFI and FI profiles.}
        \label{INJG_1D_GOW_TRANSBELL}
    \end{subfigure}
    \caption{Gas injection after water injection: pressure and saturations profiles versus cell number for the FI and the SFI methods.}
    \label{INJG_1D_GOW}
\end{figure}

This test case starts with the initial state shown in Figure \ref{INJW_1D_LARGEDT_TRANSBELL}. 
Live-gas is injected into the reservoir. The SPE~5 fluid \cite{SPE5} is used, whereas the light component is injected.
Figure \ref{INJG_1D_GOW_TRANSBELL1} shows the pressure and saturation profiles after one timestep with a CFL number of 60
obtained using the FI and the SFI methods.
The FI method required 22 iterations to converge.
One outer iteration of the SFI method consisting of three pressure and 11 compositional iterations was employed
($|R_{thermo}|_{\infty}$, $|R_{u_t}|_{\infty}$ and $|R_{\nabla \cdot u_t}|_{\infty}$ are 4.95, 0.89 and 0.57).
The SFI solutions with one outer iteration are different from the FI results.
For complete convergence, 27 iterations (three outer, 9 pressure and 18 compositional iterations) were required; 
the profiles are shown in Figure \ref{INJG_1D_GOW_TRANSBELL}.

\begin{figure}
    \centering
    \begin{subfigure}[h!]{0.8\textwidth}
        %\centering
        \includegraphics[width=\textwidth]{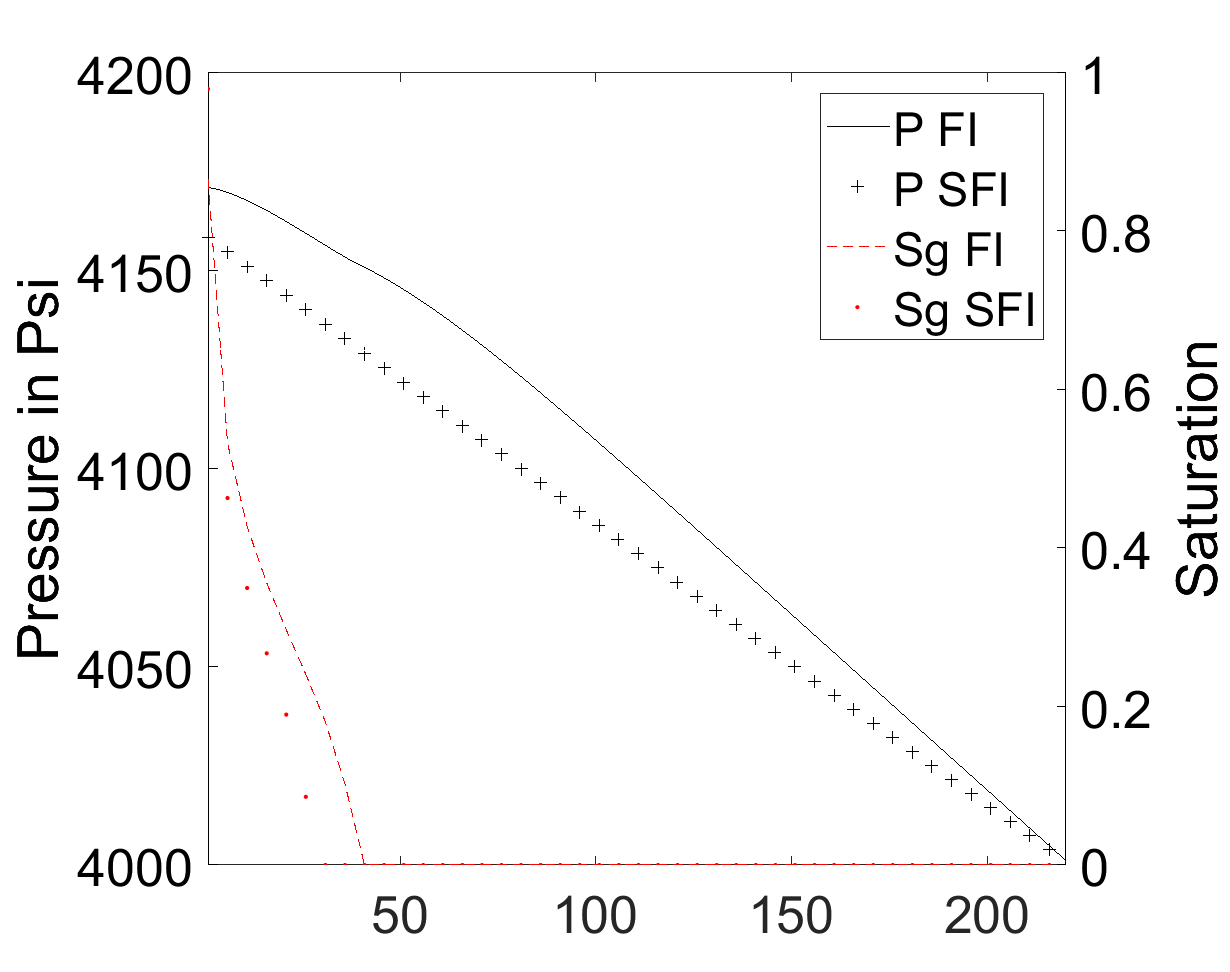}
        \caption{SFI profiles after one outer iteration and converged FI profiles.}
        \label{INJG_1D_SG00_LARGEDT200_TRANSBELL1}
    \end{subfigure}
    \\
    \begin{subfigure}[h!]{0.8\textwidth}
        %\centering
        \includegraphics[width=\textwidth]{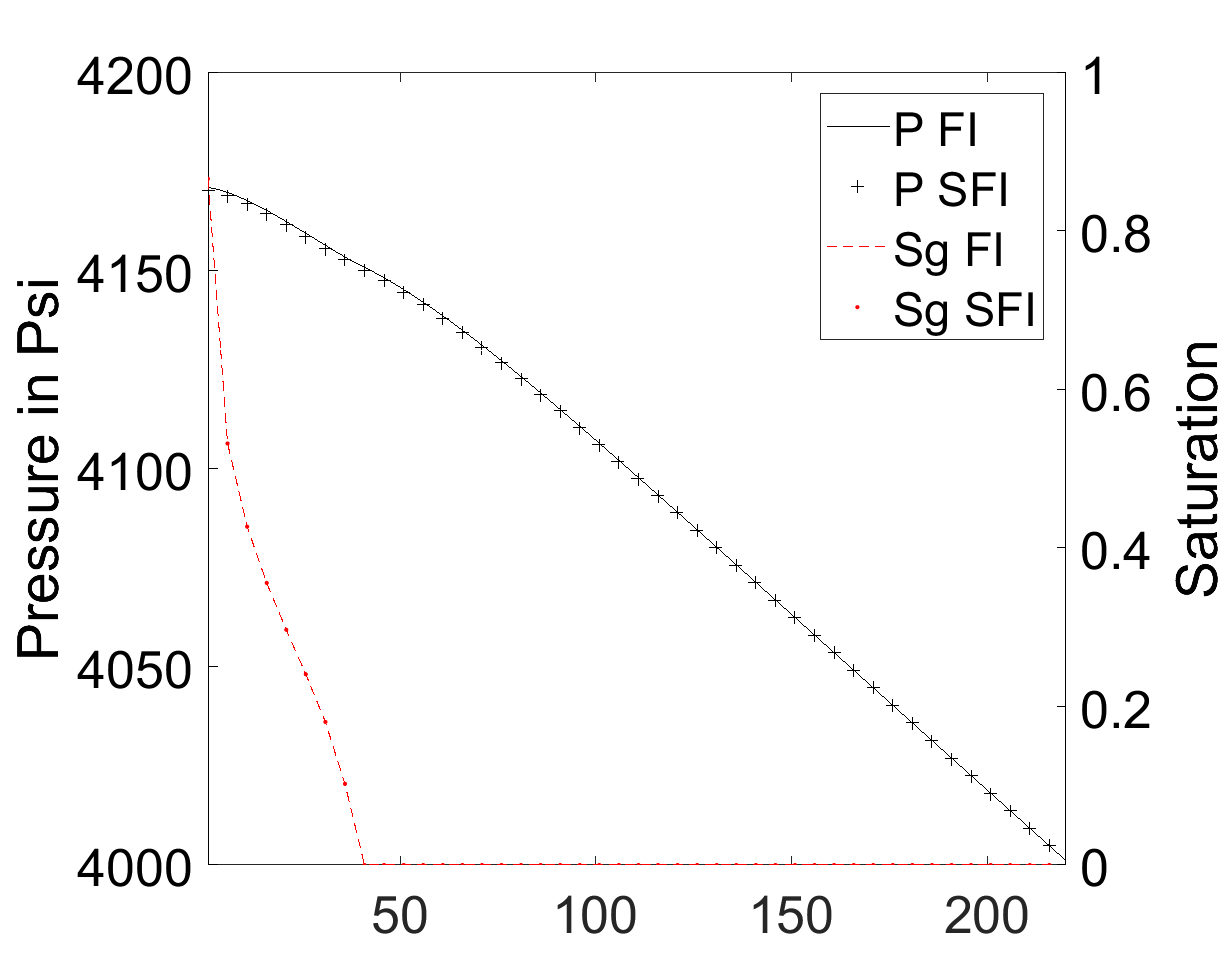}
        \caption{Converged SFI and FI profiles.}
        \label{INJG_1D_SG00_LARGEDT200_TRANSBELL}
    \end{subfigure}
    \caption{Gas injection: pressure and gas saturation profiles versus cell number for the FI and the SFI methods.}
    \label{INJG_1D_SG00_LARGEDT200}
\end{figure}

The next test case describes injection of wet-gas into a reservoir initially filled with live-oil.
Figure \ref{INJG_1D_SG00_LARGEDT200} shows pressure and gas saturation profiles after one timestep with a CFL number of 64 
using the SFI method; once after one outer iteration consisting of three pressure and 11 compositional iterations and once after
full convergence (requiring in total three outer iterations consisting of 9 pressure 
and 20 compositional iterations). The FI method requires 37 iterations to converge.
After one outer iteration of the SFI method the spiltting-errors 
$|R_{thermo}|_{\infty}$, $|R_{u_t}|_{\infty}$ and $|R_{\nabla \cdot u_t}|_{\infty}$ are 10.2, 0.90 and 0.52, respectively.

\begin{figure}
    \centering
    \begin{subfigure}[h!]{0.8\textwidth}
        %\centering
        \includegraphics[width=\textwidth]{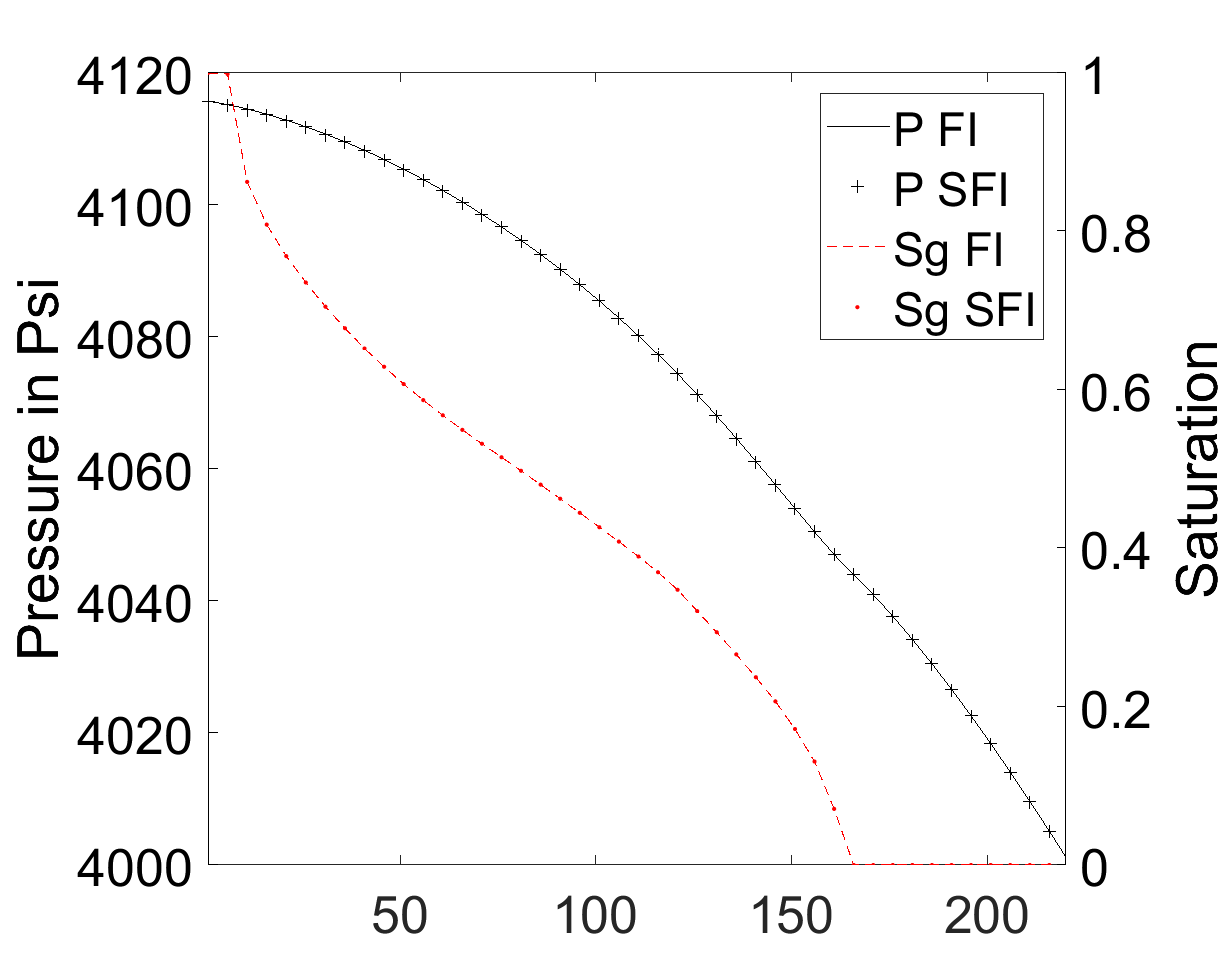}
        \caption{Initial SFI and FI profiles.}
        \label{INJG_1D_SG00_800D_LARGEDT200_START}
    \end{subfigure}
    \\
    \begin{subfigure}[h!]{0.8\textwidth}
        %\centering
        \includegraphics[width=\textwidth]{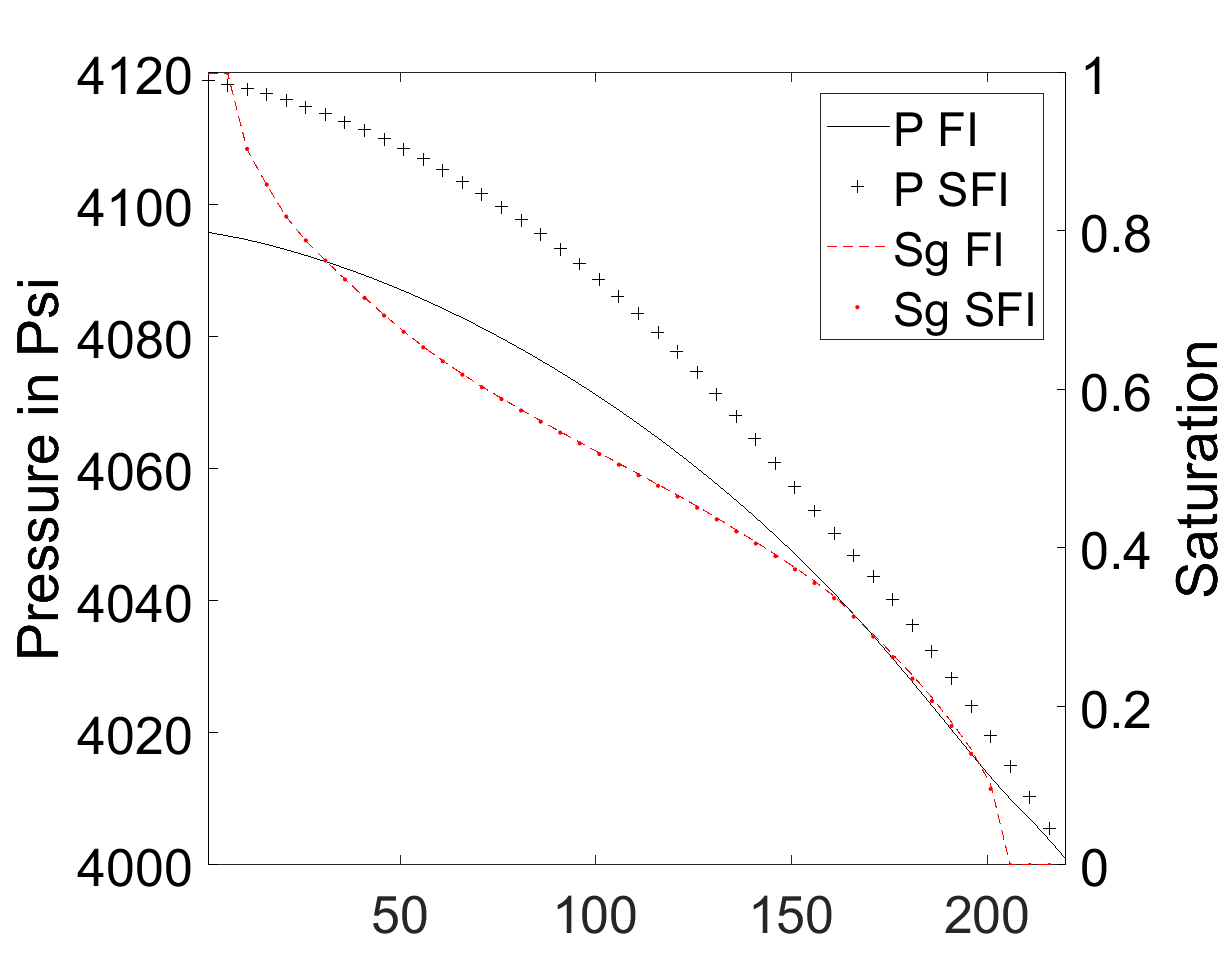}
        \caption{SFI profiles after one outer iteration and converged FI profiles for a timestep with a CFL number of 76.}
        \label{INJG_1D_SG00_800D_LARGEDT200_END}
    \end{subfigure}
    \caption{Gas injection: pressure and gas saturation profiles versus cell number for the FI and the SFI methods.}
    \label{INJG_1D_SG00_800D_LARGEDT200}
\end{figure}

\begin{figure}
    \centering
    \includegraphics[width=0.7\textwidth]{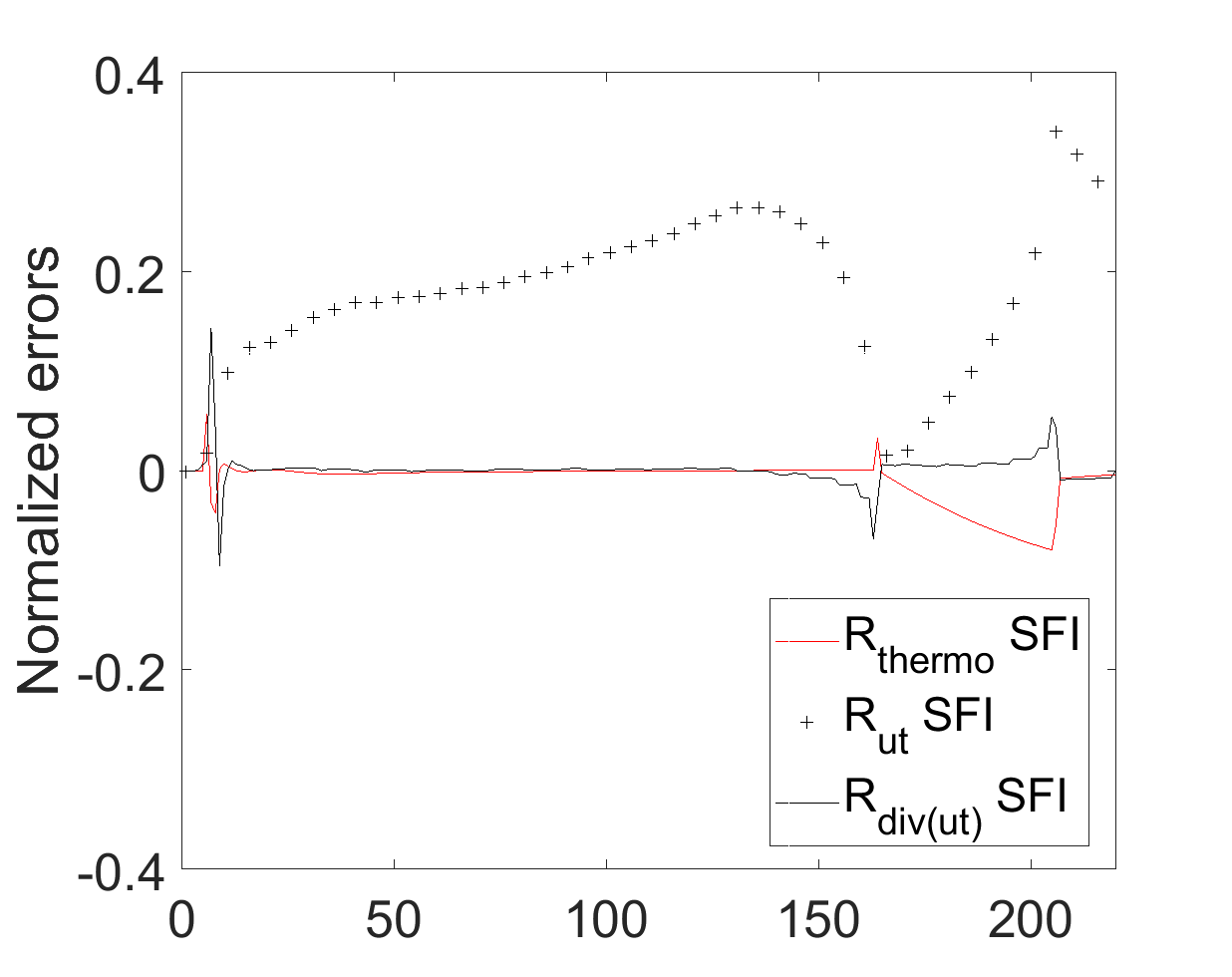}
    \caption{Gas injection: thermodynamic volume error $R_{thermo}$, total-velocity error $R_{u_t}$ and divergence of the total-velocity error $R_{\nabla \cdot u_t}$ for the SFI method after one outer iteration.}
    \label{INJG_1D_SG00_800D_LARGEDT200_ERROR}
\end{figure}

The next test case starts with the initial state of Figure \ref{INJG_1D_SG00_800D_LARGEDT200}a
and live-gas is injected into the reservoir.
Figure \ref{INJG_1D_SG00_800D_LARGEDT200}b shows pressure and gas saturation profiles 
after one timestep with a CFL number of 76
for the FI method and the SFI method stopped after one outer iteration.
The FI method requires 10 iterations to converge.
One outer iteration of the SFI method requires three pressure and 9 compositional iterations.
The saturation profiles look identical, but the pressure profiles are different.
Figure \ref{INJG_1D_SG00_800D_LARGEDT200_ERROR} shows 
the thermodynamic volume error $R_{thermo}$, 
the total-velocity error $R_{u_t}$ 
and the divergence of the total-velocity error $R_{\nabla \cdot u_t}$
profiles for the SFI method.
$|R_{thermo}|$ has values up to 8\%, $|R_{u_t}|$ has values up to 34\% and $|R_{\nabla \cdot u_t}|$ has values up to 14\%.
On the contrary to the total-velocity error $R_{u_t}$ that has a support on the whole 1D domain,
$R_{thermo}$ and $R_{\nabla \cdot u_t}$ have only non-zero values in localized areas.
This is an indication that when the compressibility and thermodynamic effects are not too strong 
we can reconstruct the correct transport in one outer iteration. 
Table \ref{INJG_1D_SG00_800D_LARGEDT200_TABLE} shows the evolution for the SFI method 
of the thermodynamic volume error, the total-velocity error, the divergence of the total-velocity error as well as 
the divergence of the total-velocity in function of the number of outer iterations.
\begin{table}[h]
\scriptsize
\centering
\caption{Gas injection: evolution of the thermodynamic volume error, the total-velocity error, the divergence of the total-velocity error and the divergence of the total-velocity in function of the number of outer iterations.}
\label{INJG_1D_SG00_800D_LARGEDT200_TABLE}
\begin{tabular}{|c|c|c|c|c|c|c|}
\hline
Outer iterations & P iterations & C iterations & $|R_{thermo}|_{\infty}$ & $|R_{u_t}|_{\infty}$ & $|R_{\nabla \cdot u_t}|_{\infty}$ & $|\overline{\nabla \cdot  u_t}|_{\infty}$ \\
\hline
1 & 3 & 9 & 0.08 & 0.34 & 0.14 & 0.14\\
2 & 2 & 2 & 0.02 & 0.02 & 0.02 & 0.02 \\
3 & 1 & 1 & 1e-3 & 2e-4 & 3e-4 & 3e-3 \\
4 & 1 & 1 & 3e-5 & 7e-6 & 6e-6 & 3e-3\\
5 & 1 & 1 & 2e-6 & 1e-6 & 2e-6 & 3e-3 \\
6 & 1 & 1 & 2e-8 & 4e-8 & 4e-8 & 3e-3 \\
\hline
\end{tabular}
\end{table}
We observe that the splitting errors $|R_{thermo}|_{\infty}$, $|R_{u_t}|_{\infty}$ and $|R_{\nabla \cdot u_t}|_{\infty}$ decrease with the outer iterations. 
At some point, the divergence of the total-velocity, $\overline{\nabla \cdot  u_t}$, converges to the true value, since the flow is compressible.

% Profile plot

\begin{figure}
    \centering
    \begin{subfigure}[h!]{0.8\textwidth}
        %\centering
        \includegraphics[width=\textwidth]{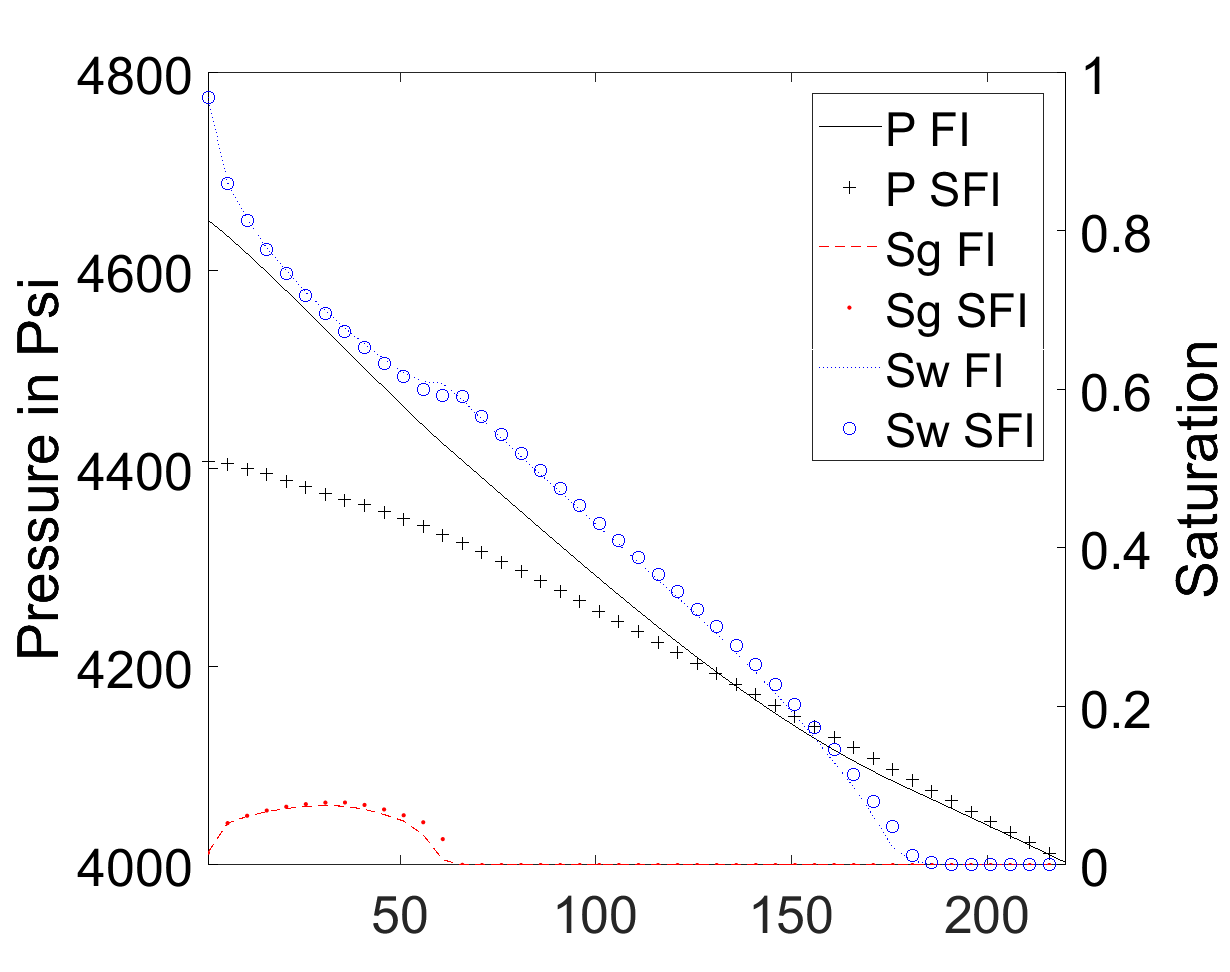}
        \caption{SFI profiles after one outer iteration and converged FI profiles.}
        \label{INJW_1D_GOW_TRANSBELL1}
    \end{subfigure}
    \\
    \begin{subfigure}[h!]{0.8\textwidth}
        %\centering
        \includegraphics[width=\textwidth]{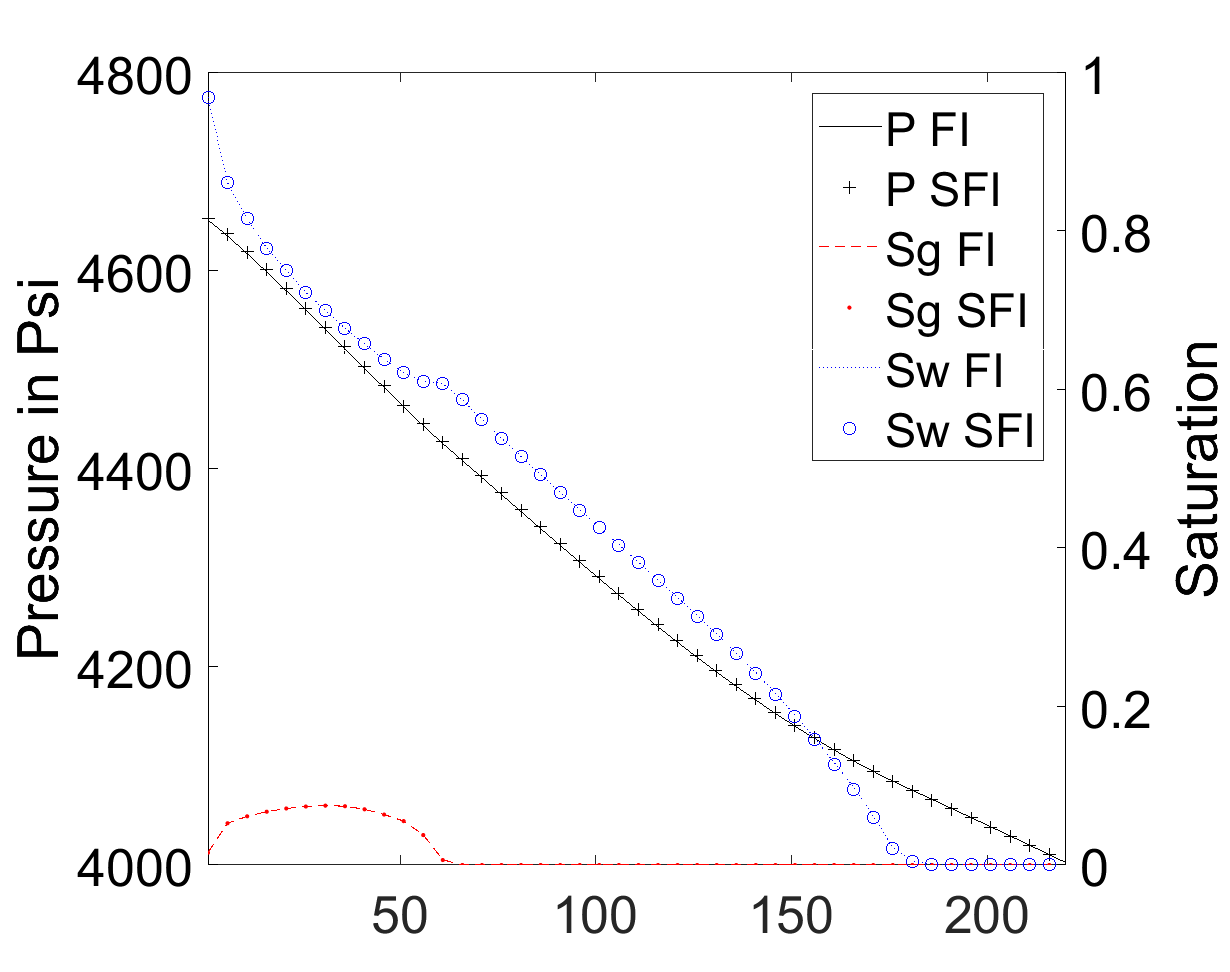}
        \caption{Converged SFI and FI profiles.}
        \label{INJW_1D_GOW_TRANSBELL}
    \end{subfigure}
    \caption{Water injection: pressure and saturations profiles versus cell number for the FI and the SFI methods.}
    \label{INJW_1D_GOW}
\end{figure}

Starting from the final state of Figure \ref{INJG_1D_SG00_LARGEDT200}, water is injected for one timestep with a CFL number of 190.
Figure \ref{INJW_1D_GOW} shows the pressure and saturations profiles after this timestep.
The FI method required 18 iterations to converge.
One outer iteration of the SFI method required three pressure and 16 compositional iterations.
After this outer iteration, the maximum values of $|R_{thermo}|_{\infty}$, $|R_{u_t}|_{\infty}$ and $|R_{\nabla \cdot u_t}|_{\infty}$
%HAT: are you sure about cell 1
were 0.34 (in cell 39), 0.89 (in cell 1) and 0.05 (in cell 1), respectively.
The pressure and gas-saturation profiles obtained after one iteration with  the SFI and FI methods are different. 
To achieve convergence, SFI scheme required two outer iterations consisting of a total of 6 pressure and 18 compositional iterations.

Similar test cases have been simulated with the "system 1" fluid from \cite{Zaydullin:2012} 
and the decane/CO$_2$/methane fluid from \cite{MoynerRSC:2017}, 
and the same observations apply.

\subsection{2D test cases}

In this section, we present 2D test cases with gas and water injection.

\subsubsection{2D Case-1: Gas and Water Injection}

\begin{figure}
    \centering
    \begin{subfigure}[h!]{0.9\textwidth}
        %\centering
        \includegraphics[width=\textwidth]{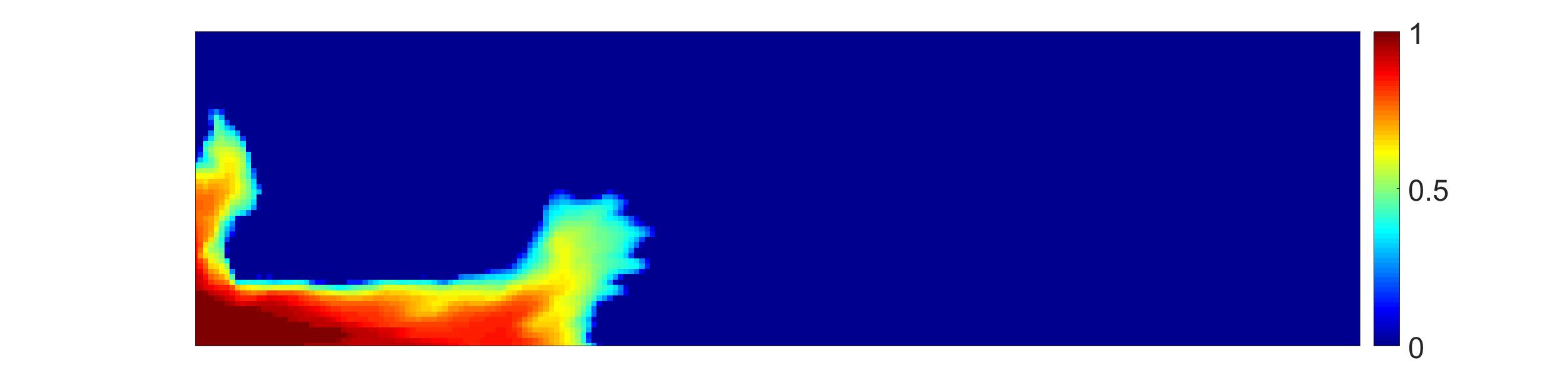}
        \caption{$S_g$.}
        \label{images_IPAM1:IPAM1_4000_4200_TRANSBELL1P4C8FS_SGAS_RP9}
    \end{subfigure}
    \\
    \begin{subfigure}[h!]{0.9\textwidth}
        %\centering
        \includegraphics[width=\textwidth]{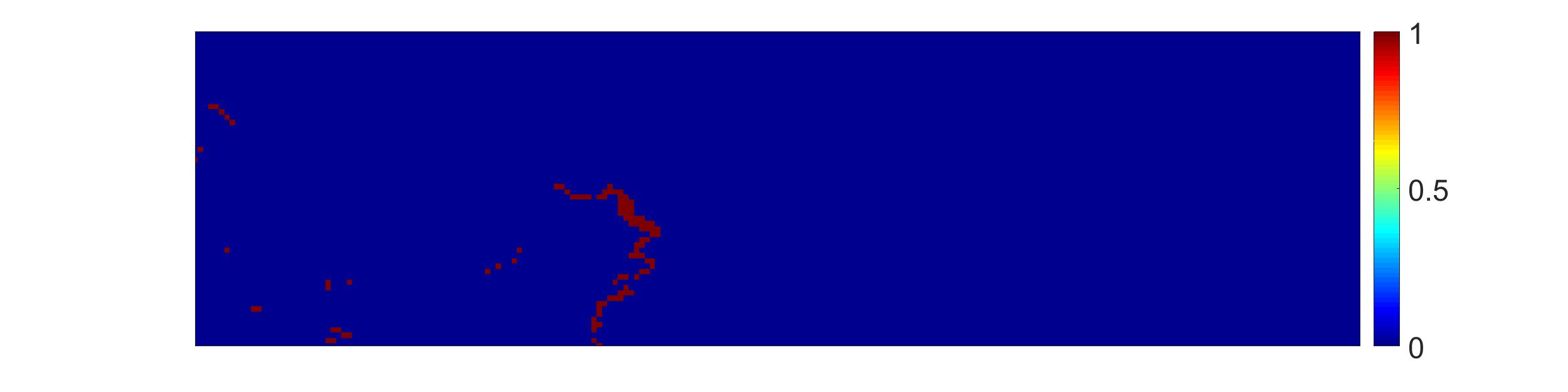}
        \caption{Volume error above 1\%. Error ranges from -7\% to +7\%.}
        \label{images_IPAM1:IPAM1_4000_4200_TRANSBELL1P4C8FS_VOLERROR_0P01_RP9}
    \end{subfigure}
    \\
    \begin{subfigure}[h!]{0.9\textwidth}
        %\centering
        \includegraphics[width=\textwidth]{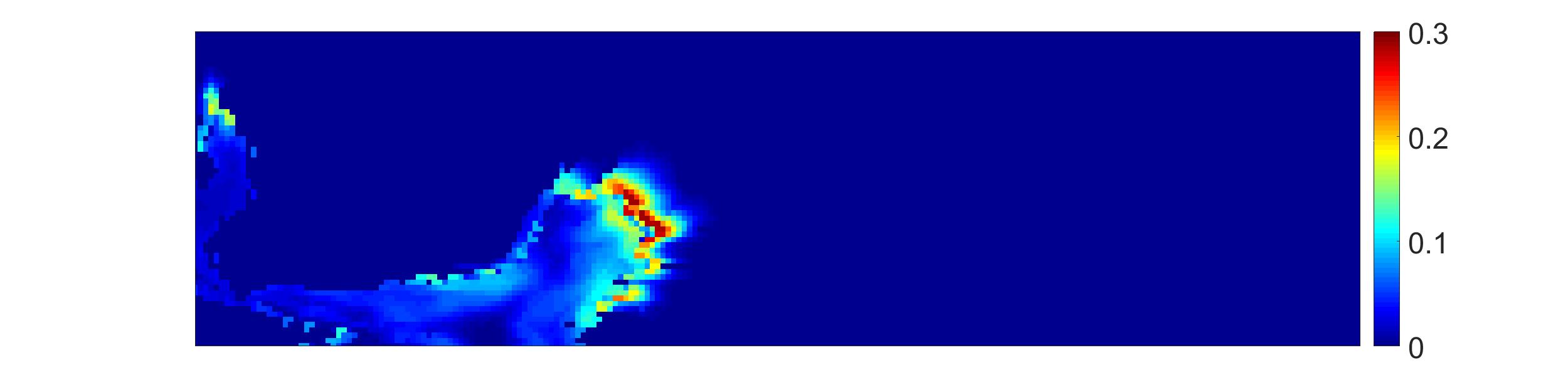}
        \caption{Total-velocity error reaches up to 30\%.}
        \label{images_IPAM1:IPAM1_4000_4200_TRANSBELL1P4C8FS_QTERROR_RP9}
    \end{subfigure}
    \\
    \begin{subfigure}[h!]{0.9\textwidth}
        %\centering
        \includegraphics[width=\textwidth]{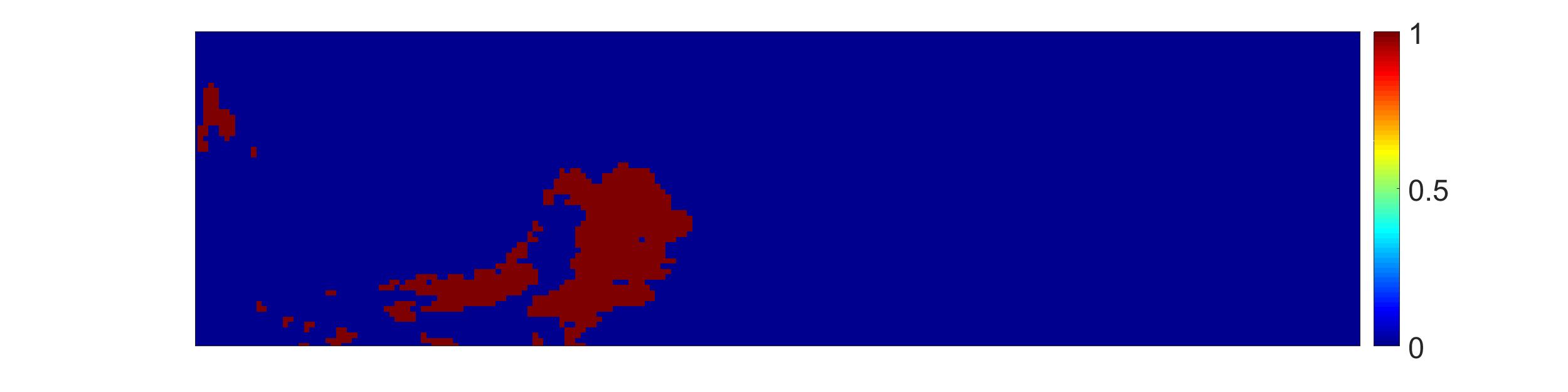}
        \caption{Velocity error exceeding 5\%.}
        \label{images_IPAM1:IPAM1_4000_4200_TRANSBELL1P4C8FS_QTERROR_0P05_RP9}
    \end{subfigure}
    \\
    \begin{subfigure}[h!]{0.9\textwidth}
        %\centering
        \includegraphics[width=\textwidth]{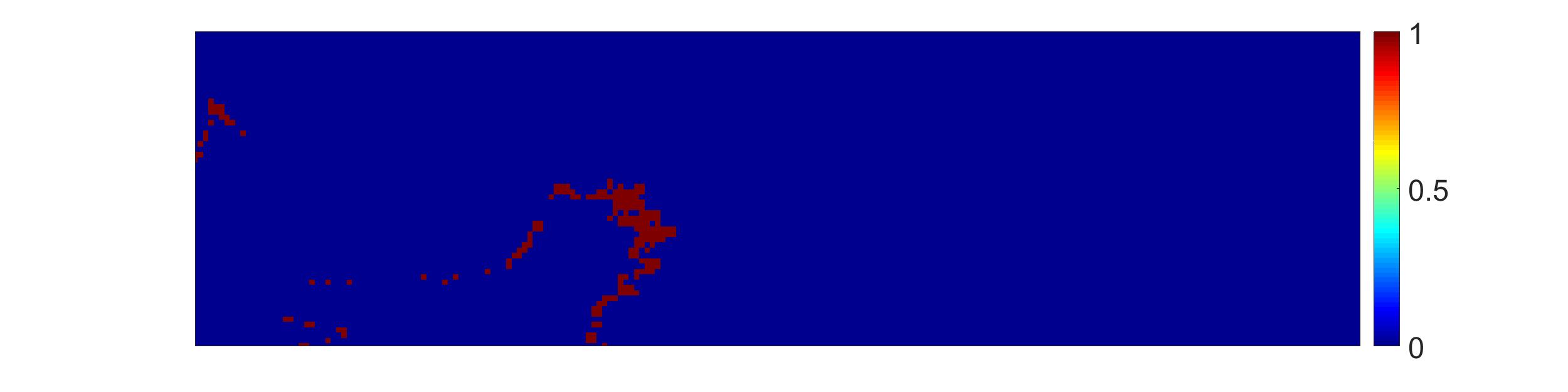}
        \caption{Divergence of velocity error exceeding 5\%. Error ranges from -39\% to +44\%.}
        \label{images_IPAM1:IPAM1_4000_4200_TRANSBELL1P4C8FS_QTDIV_0P05_RP9}
    \end{subfigure}
    \caption{Maps for 2D Case-1 after application of one outer iteration of the sequential algorithm.}
    \label{IPAM1_SEQ}
\end{figure}

The first test case is from Moncorg\'e et al. \cite{Moncorge:2017}.
The light component is injected into multi-component oil at the lower-left corner of the model; the producer, 
for which the bottom-hole pressure (BHP) is set to 4000 Psi, is located at the upper-right corner of the model. See
figure \ref{images_IPAM1:IPAM1_4000_4200_TRANSBELL1P4C8FS_SGAS_RP9}.
We study the behaviors for one particular timestep.
Figure \ref{IPAM1_SEQ} shows the solution of the SFI method after one outer iteration.
The mass balance equations are converged, but the residuals $R_{thermo}$ and $R_{u_t}$ are `relaxed'.
The CFL numbers for explicit treatment of saturation (CFL$_S$) and compositions (CFL$_X$) \cite{Coats:2003CFL} 
are 33 and 224 for the timestep, respectively.
The FI method converges in 6 iterations.
For the SFI method, we summarize in table \ref{IPAM1_4000_4200} the evolution of the errors in the thermodynamic volume, total-velocity, and divergence of the total-velocity, as functions of the number of outer iterations.
We observe that the splitting errors decrease with the outer iterations. 
Since the flow is compressible,  the divergence of the total-velocity $\overline{\nabla \cdot  u_t}$ converges to the correct nonzero value.
 
\begin{table}[h]
\scriptsize
\centering
\caption{Evolution of the thermodynamic volume error, the total-velocity error, the divergence of the total-velocity error and the divergence of the total-velocity as functions of the number of outer iterations for 2D Case-1.}
\label{IPAM1_4000_4200}
\begin{tabular}{|c|c|c|c|c|c|c|}
\hline
Outer iterations & P iterations & C iterations & $|R_{thermo}|_{\infty}$ & $|R_{u_t}|_{\infty}$  & $|R_{\nabla \cdot  u_t}|_{\infty}$ & $|\overline{\nabla \cdot  u_t}|_{\infty}$ \\
\hline
1 & 2 & 6 & 0.07 & 0.29 & 0.44 & 0.44\\
2 & 2 & 3 & 9e-3 & 7e-2 & 0.12 & 0.18 \\
3 & 2 & 1 & 1e-3 & 7e-3 & 9e-3 & 8e-2\\
4 & 1 & 1 & 9e-5 & 1e-3 & 1e-3 & 8e-2\\
5 & 1 & 1 & 6e-6 & 1e-4 & 1e-4 & 8e-2\\
6 & 1 & 1 & 6e-7 & 1e-5 & 1e-5 & 8e-2\\
\hline
\end{tabular}
\end{table}

After one outer iteration with two pressure and six compositional solves, 
we observe highly localized errors at the front between the gas and the oil phases in the thermodynamic volume equation with values that amount up to 7\%. 
The cells with errors of more than 1\% for the volume are plotted in Figure \ref{images_IPAM1:IPAM1_4000_4200_TRANSBELL1P4C8FS_VOLERROR_0P01_RP9}.
Total-velocity errors as large as 29\% are observed, and they have a large support as seen in Figure \ref{images_IPAM1:IPAM1_4000_4200_TRANSBELL1P4C8FS_QTERROR_RP9}.
The cells with errors of more than 5\% in the total-velocity are plotted in Figure \ref{images_IPAM1:IPAM1_4000_4200_TRANSBELL1P4C8FS_QTERROR_0P05_RP9}.
Figure \ref{images_IPAM1:IPAM1_4000_4200_TRANSBELL1P4C8FS_QTDIV_0P05_RP9}
shows that errors greater than 5\% in the total-velocity divergence are localized.
Nevertheless, the largest error is about 44\%.
For this timestep, we required three outer iterations (a total of six  pressure and 10 compositional solves) to converge to the FI solution.

\begin{figure}
    \centering
    \begin{subfigure}[h!]{0.9\textwidth}
        %\centering
        \includegraphics[width=\textwidth]{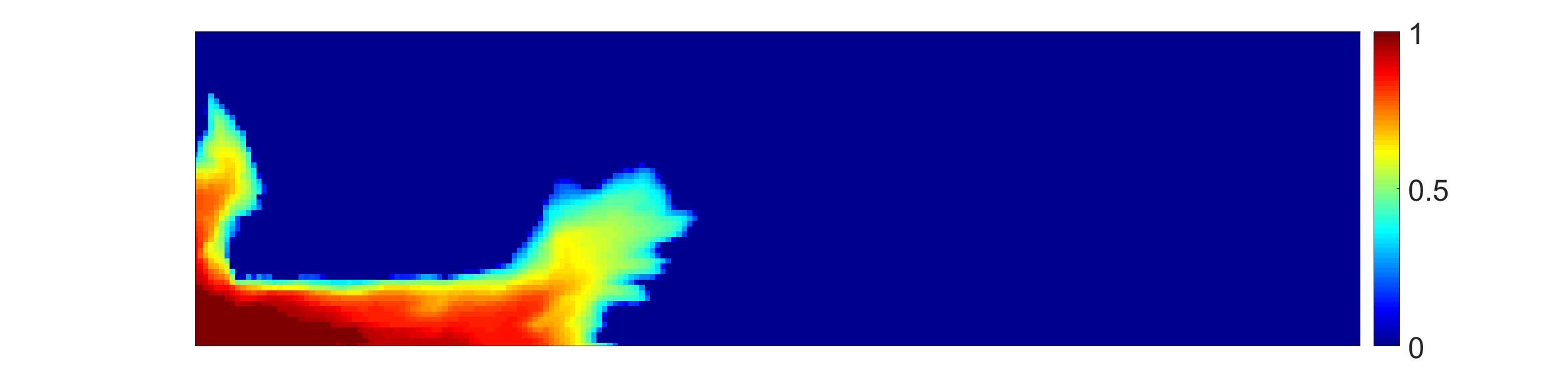}
        \caption{After gas injection at the end of the first period.}
        \label{images_IPAM1_1}
    \end{subfigure}
    \\
    \begin{subfigure}[h!]{0.9\textwidth}
        %\centering
        \includegraphics[width=\textwidth]{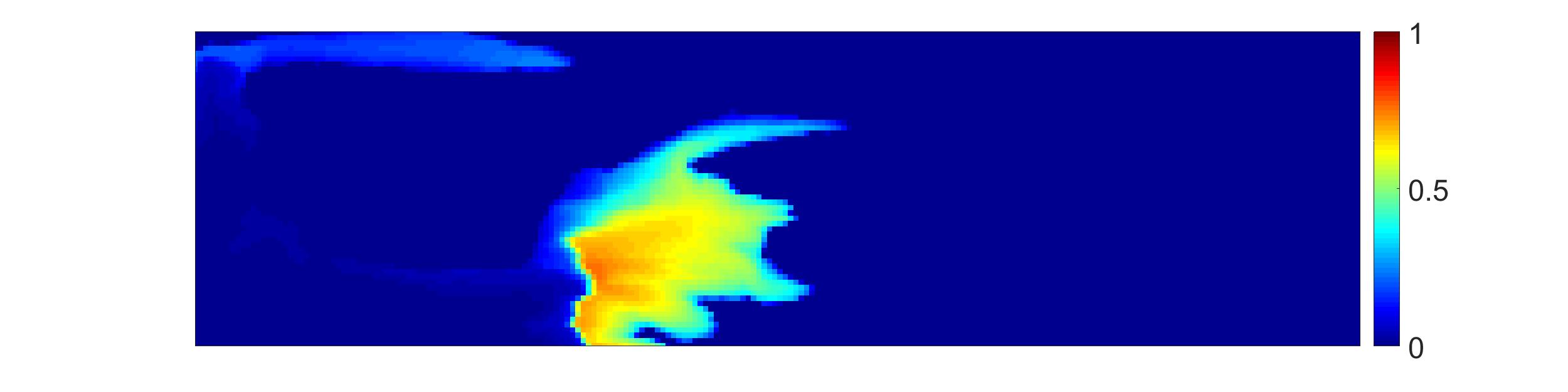}
        \caption{After water injection at the end of the second period.}
        \label{images_IPAM1_2}
    \end{subfigure}
    \\
    \begin{subfigure}[h!]{0.9\textwidth}
        %\centering
        \includegraphics[width=\textwidth]{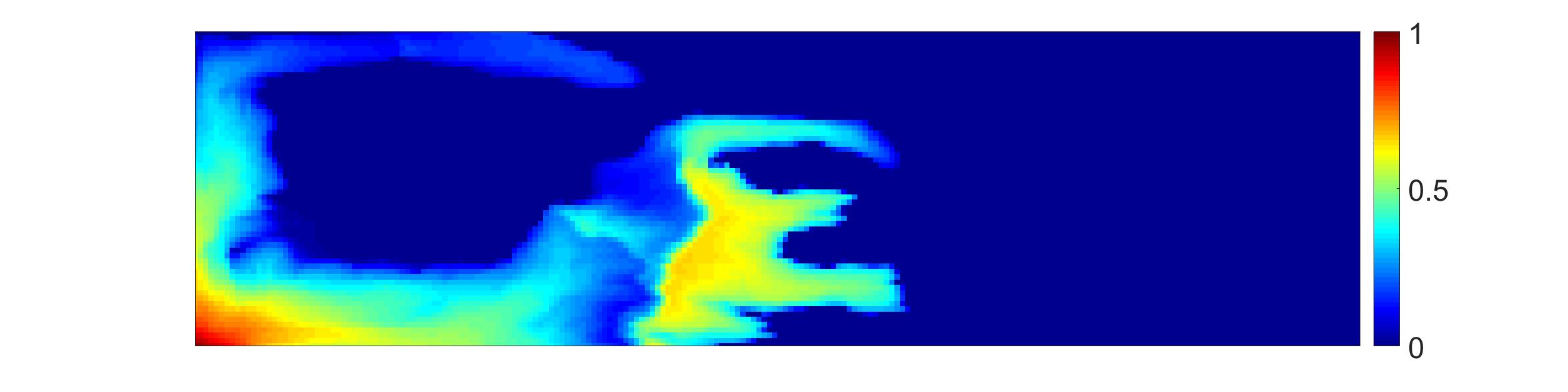}
        \caption{After gas injection at the end of the third period.}
        \label{images_IPAM1_3}
    \end{subfigure}
    \caption{Gas saturation for 2D Case-1.}
    \label{CASE-1-SAT}
\end{figure}

\begin{figure}
    \centering
    \includegraphics[width=0.7\textwidth]{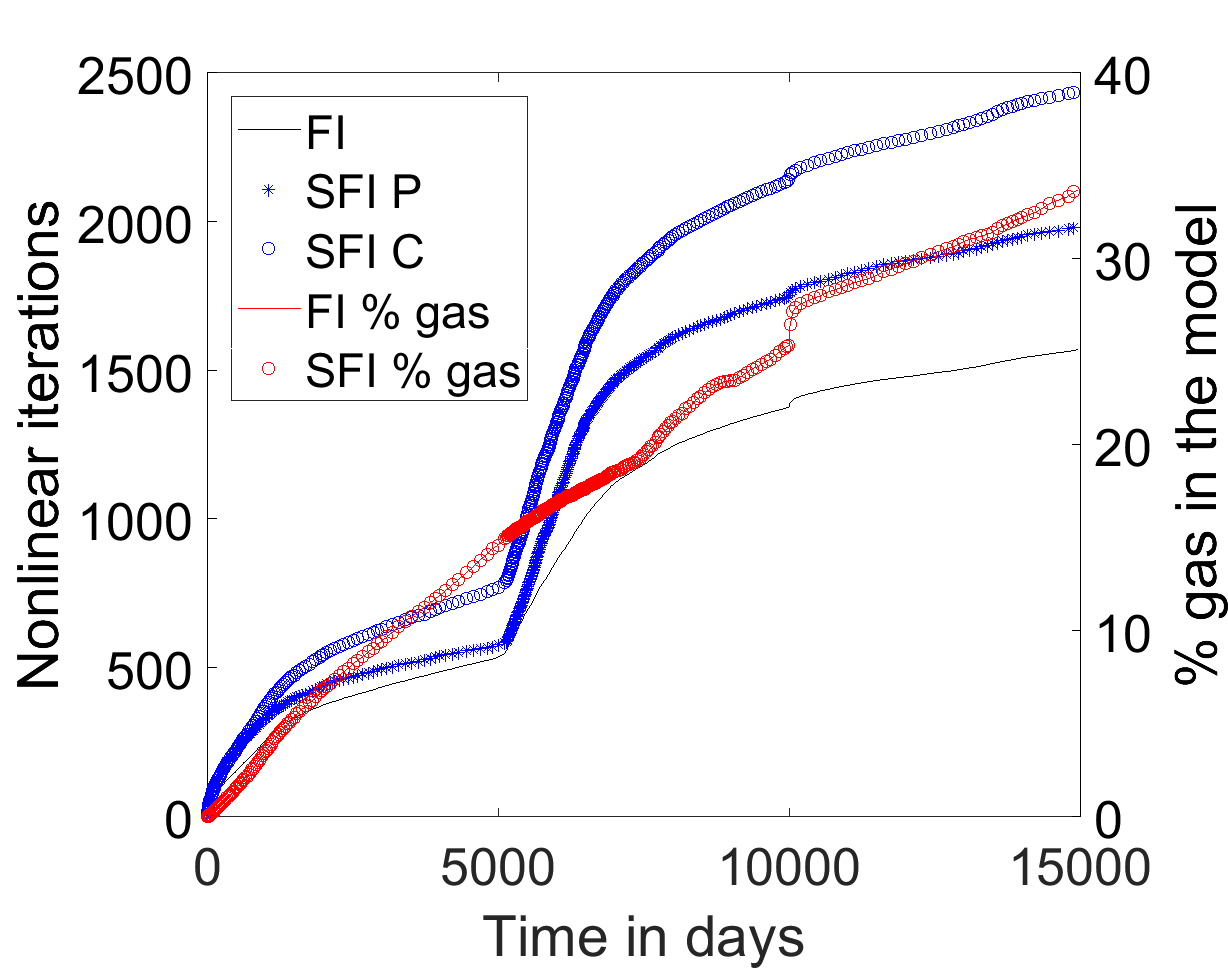}
    \caption{Cumulative Newton iterations for the FI method and cumulative pressure and composition iterations for the SFI method for 2D Case-1. The percentage of gas in the model for the FI and SFI methods is also shown.}
    \label{CASE-1-ITER}
\end{figure}

\begin{figure}
    \centering
    \includegraphics[width=0.7\textwidth]{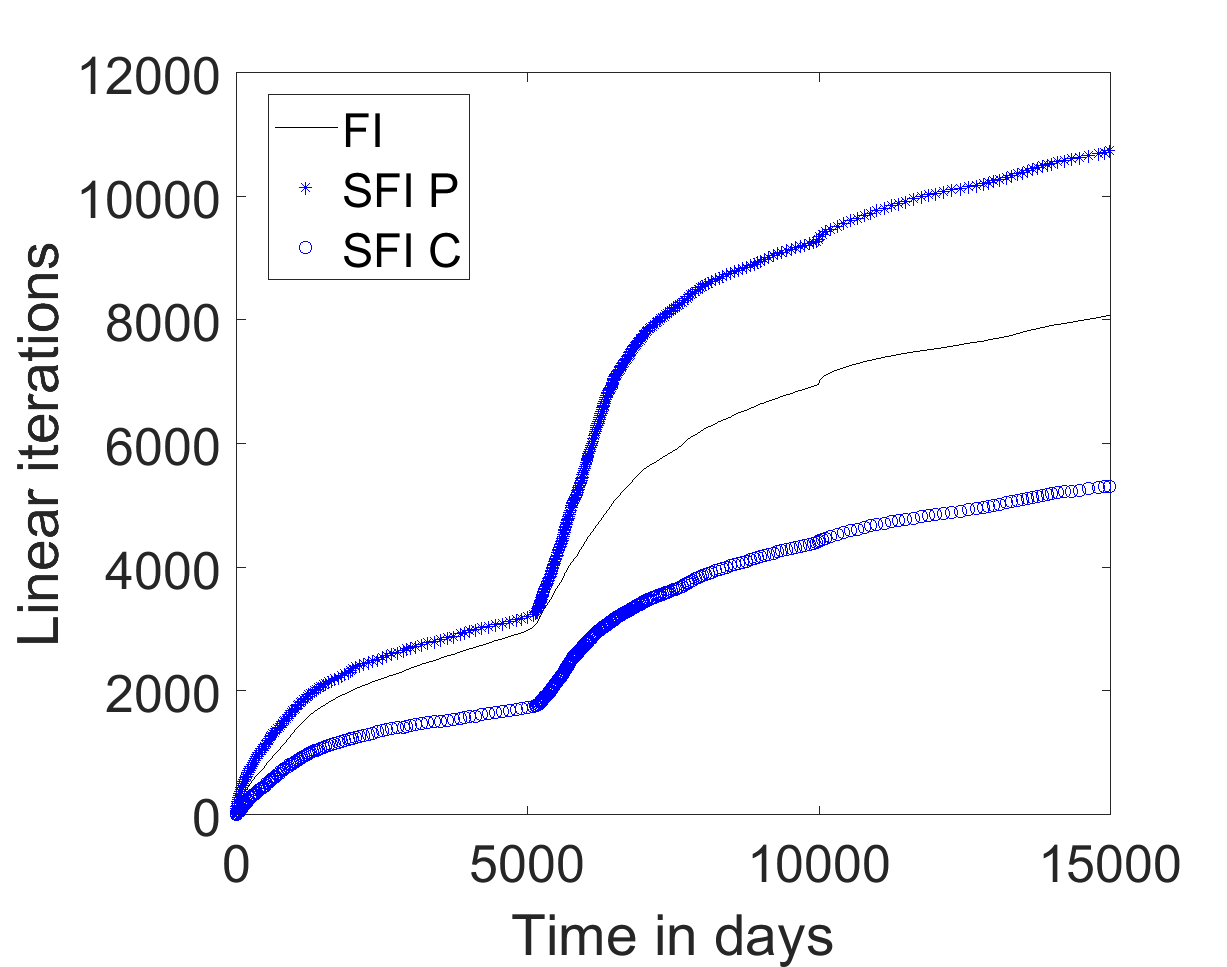}
    \caption{Cumulative CPR iterations for the systems of the FI method and the pressure of the SFI method and cumulative ILU0-GMRES iterations for the composition systems of the SFI method for 2D Case-1.}
    \label{CASE-1-LINS}
\end{figure}

We now study the behaviors for the full simulation. The first period involves injection of the light-component at a rate of six MSCF/Day;
the second period entails water injection at a rate of four STB/Day, 
and the third period involves light-component injection at a rate of six MSCF/Day.
This injection strategy leads to complex behaviors with gas fronts chased by water fronts that force the gas to dissolve back into the oil phase.
FI requires 399 and SFI 401 timesteps to reach convergence.
Figures \ref{images_IPAM1_1}, \ref{images_IPAM1_2} and \ref{images_IPAM1_3} show the gas saturation profiles 
at the end of each of the three periods.
For the first period, the average CFL numbers for the explicit treatment of saturation (CFL$_S$) and for the explicit treatment
of compositions (CFL$_X$) are 40 and 71, respectively, and the corresponding maximum values are 204 and 228.
For the second period, the average values of CFL$_S$ and CFL$_X$ are 39 and 15, respectively, and their maximum values are 500 and 136.
For the third period, the average values of CFL$_S$ and CFL$_X$ are 182 and 191, respectively, and the maximum values are 635 and 490.
Figure \ref{CASE-1-ITER} shows the cumulative number of Newton iterations for the FI method and the cumulative number of 
pressure and composition iterations for the SFI method, as well as the 
percentage of gas in the model for both the FI and SFI methods.
For this test case, the SFI method requires 26\% more pressure iterations and 55\% more composition iterations than the FI method Newton iterations.
The percentage of gas is exactly the same for both models.
Figure \ref{CASE-1-LINS} shows the cumulative number of CPR(AMG,ILU0)-FGMRES iterations required by the FI method and
by the pressure solve of the SFI method and the cumulative number of ILU0-GMRES iterations required by the SFI method for the compositional system.
Both the full system of the FI method and the pressure systems of the SFI method require around 5 CPR iterations to converge, 
but the compositional systems of the SFI method only require 2.2 ILU0-GMRES iterations to converge.
As a result, for this test case the SFI method takes 33\% more linear pressure iterations than the FI method, 
but only 65\% of the FI linear composition iterations.
This test case is quite challenging for SFI methods, as the coupling between flow and transport is strong.


\subsubsection{2D Case-2: Gas and Water Injection}
\begin{figure}
    \centering
    \begin{subfigure}[h!]{0.9\textwidth}
        %\centering
        \includegraphics[width=\textwidth]{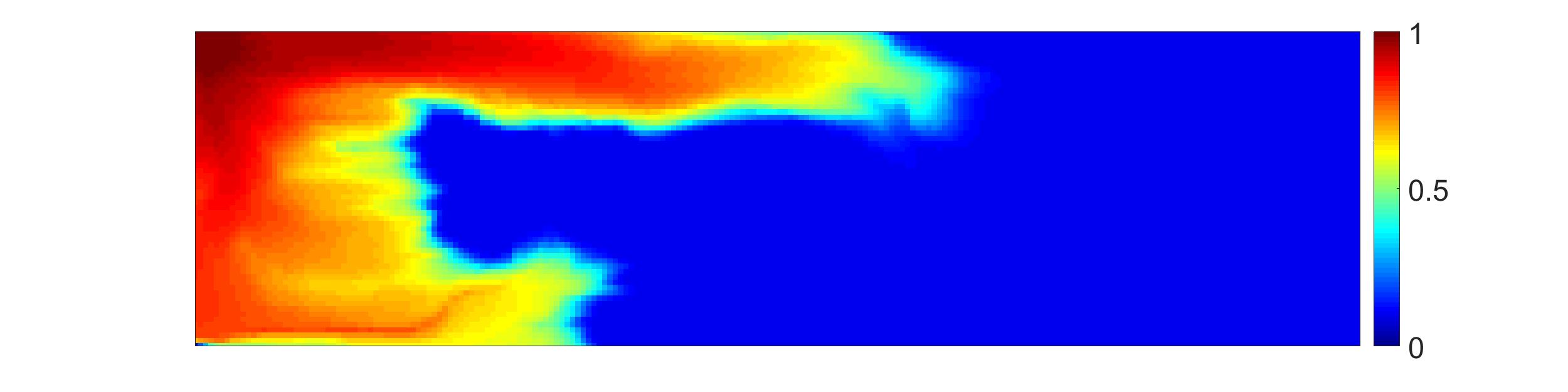}
        \caption{Water saturation at the end of the simulation.}
        \label{images_IPAM2_SW}
    \end{subfigure}
    \\
    \begin{subfigure}[h!]{0.9\textwidth}
        %\centering
        \includegraphics[width=\textwidth]{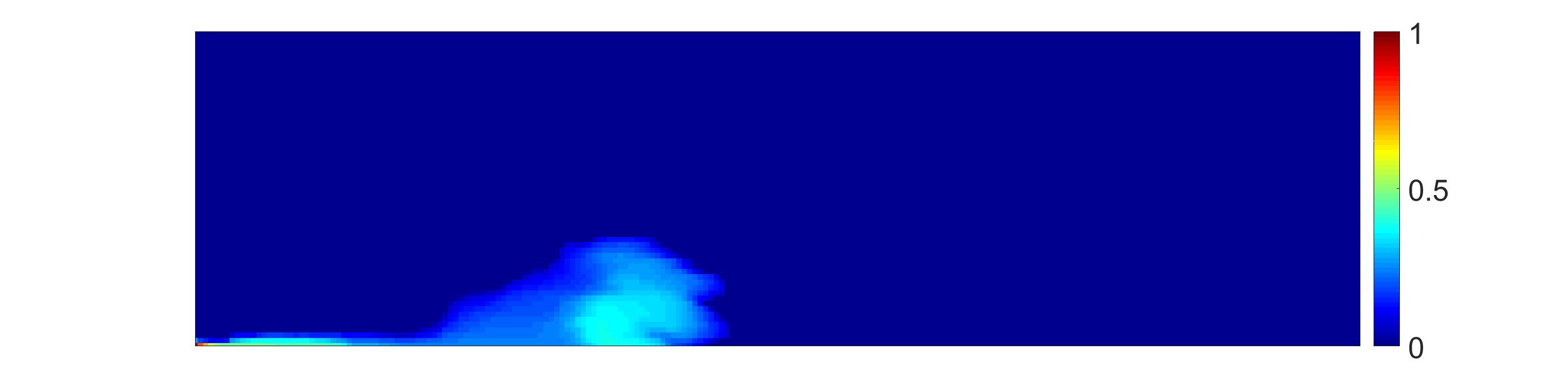}
        \caption{Gas saturation at the end of the simulation.}
        \label{images_IPAM2_SG}
    \end{subfigure}
    \caption{Saturation maps for 2D Case-2.}
    \label{CASE-2-SAT}
\end{figure}

\begin{figure}
    \centering
    \includegraphics[width=0.7\textwidth]{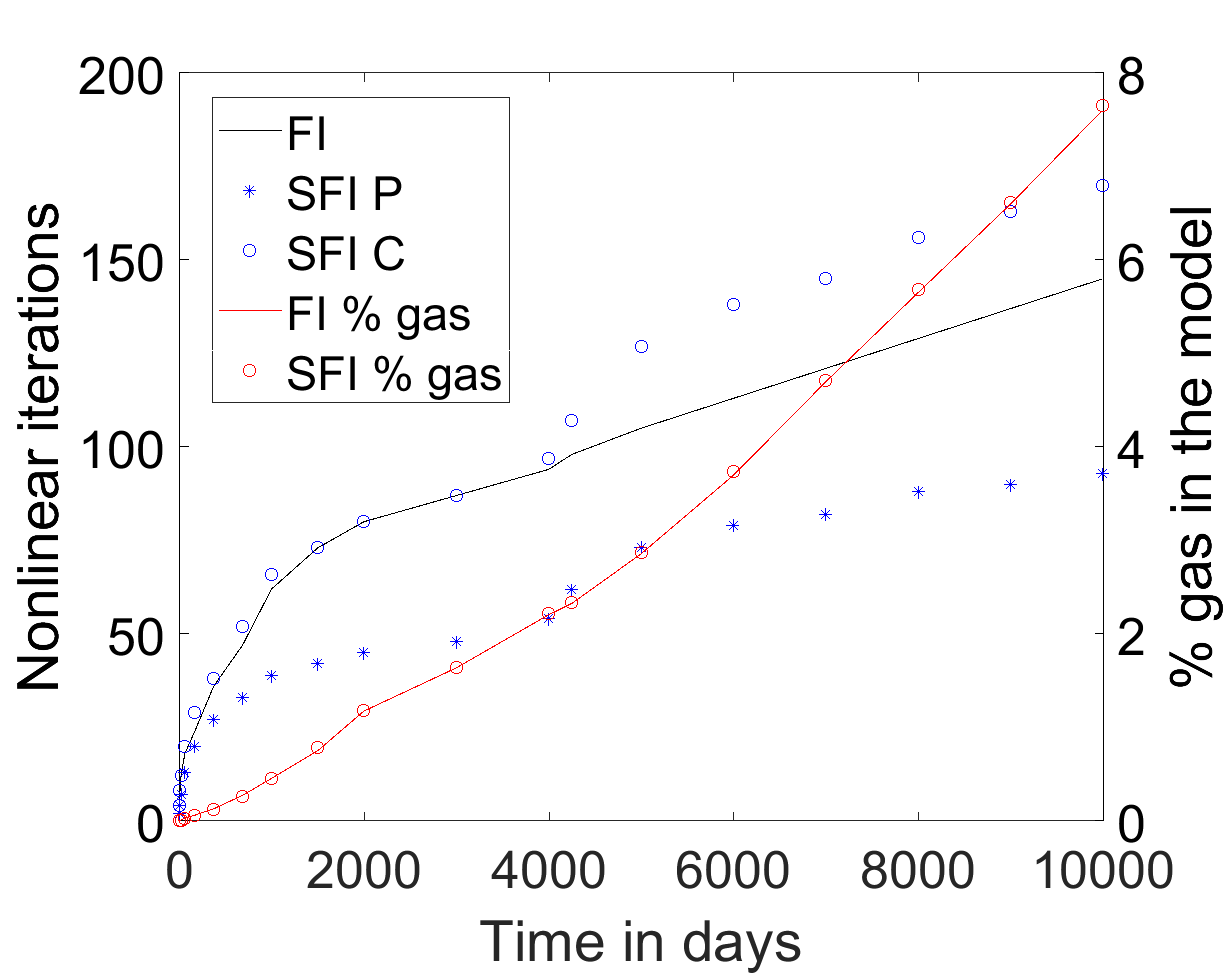}
    \caption{Cumulative Newton iterations for the FI method and cumulative pressure and composition iterations for the SFI method for 2D Case-2. The percentage of gas in the model for the FI and SFI methods is also shown.}
    \label{CASE-2-ITER}
\end{figure}

\begin{figure}
    \centering
    \includegraphics[width=0.7\textwidth]{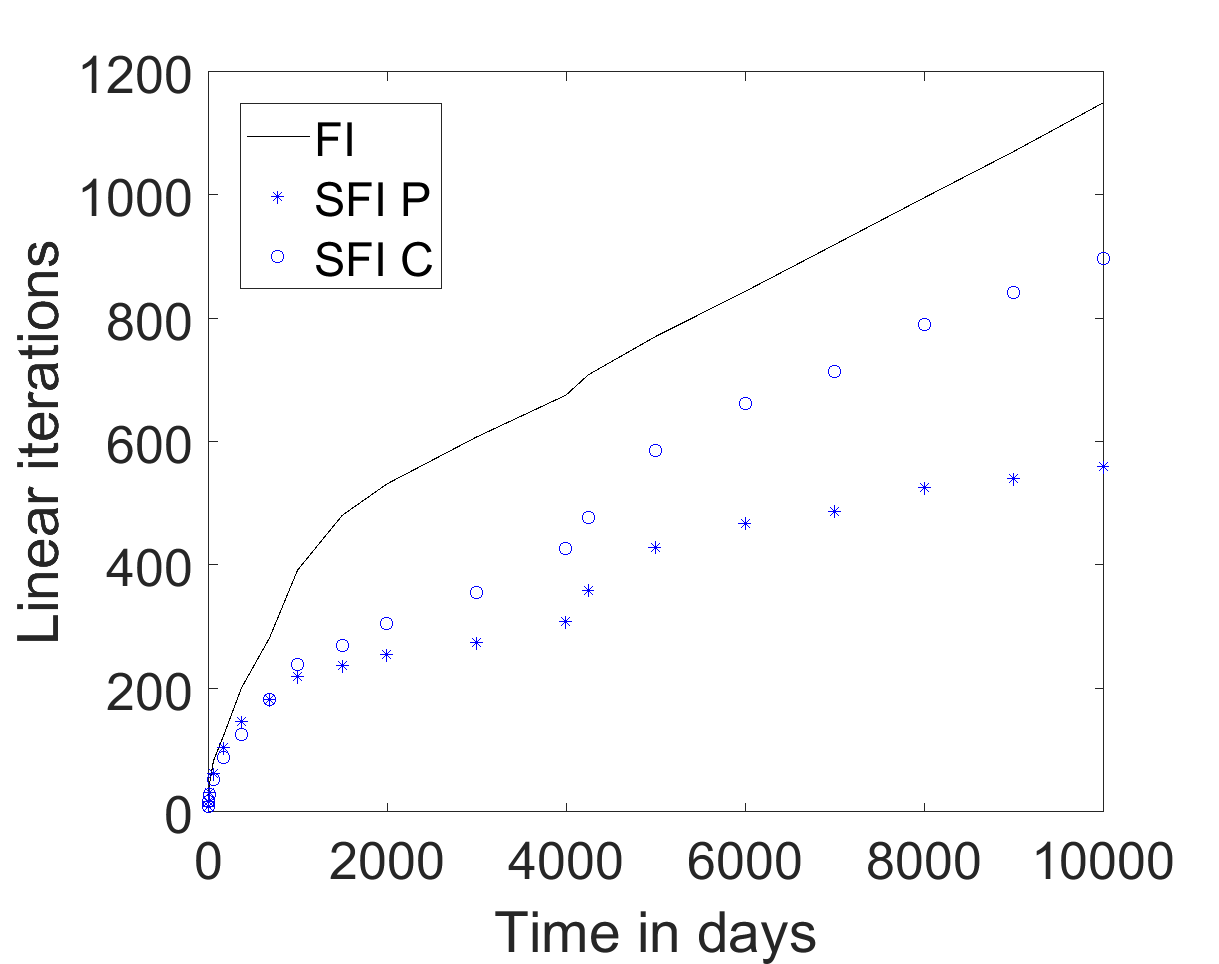}
    \caption{Cumulative CPR iterations for the systems of the FI method and the pressure of the SFI method and cumulative ILU0-GMRES iterations for the composition system of the SFI method for 2D Case-2.}
    \label{CASE-2-LINS}
\end{figure}

% HAT-2-old the numbering is confusing
The second test case is challenging for the FI method, but easier for the SFI method.
Water is injected at the top-left corner at a rate of four STB/Day and gas at the bottom-left corner at a rate of one MSCF/Day.
Production occurs at the top-right and bottom-right corners, where the BHP is set to 4000 Psi. To increase
the velocity of the water phase, the porosity in the 20 top layers (out of 60 layers) is reduced by a factor
of 100. Note that such contrasts are realistic and represent low porosity corridors. 
Both the FI and SFI methods require 19 timesteps to converge.
Figures \ref{images_IPAM2_SW} and \ref{images_IPAM2_SG} show the water and gas saturation maps at the end of the simulation.
A large part of the reservoir has been flooded by water and experiences very fast velocities; gas is
only present in the lower part of the reservoir. While there is no gas-phase in the reservoir initially, at
the end of the simulation almost 8\% of the cells have gas. 
The average CFL number per timestep is 480 and the corresponding maximum value is 1140.
Figure \ref{CASE-2-ITER} shows the cumulative number of Newton iterations for the FI method and the cumulative number of 
pressure and composition iterations for the SFI method as well as the 
percentage of gas in the model for both the FI and SFI methods.
For this case, the SFI method requires 36\% fewer pressure iterations and 17\% more composition iterations than the FI method Newton iterations.
The percentage of gas for the models is exactly the same.
Figure \ref{CASE-2-LINS}, shows the cumulative number of CPR(AMG,ILU0)-FGMRES iterations for the FI method and
the pressure solves of the SFI method and the cumulative number of ILU0-GMRES iterations for the compositional system of the SFI method.
The full system of the FI method requires in average around 8 CPR iterations to converge,
while the pressure systems of the SFI method requires around 6 CPR iterations to converge.
The compositional systems of the SFI method requires in average 5.3 ILU0-GMRES iterations to converge.
As a result, the SFI method for this test case takes 49\% of the linear pressure iterations and 78\% of the linear composition iterations of the FI method.
The strong oil-water flow is challenging for the FI method and very well handled by the SFI method,
and the mild compositional effects are handled well by both methods.
This is a configuration that is closer to real reservoir simulation scenarios.


\subsection{3D test cases}

\begin{figure}
    \centering
    \includegraphics[width=0.7\textwidth]{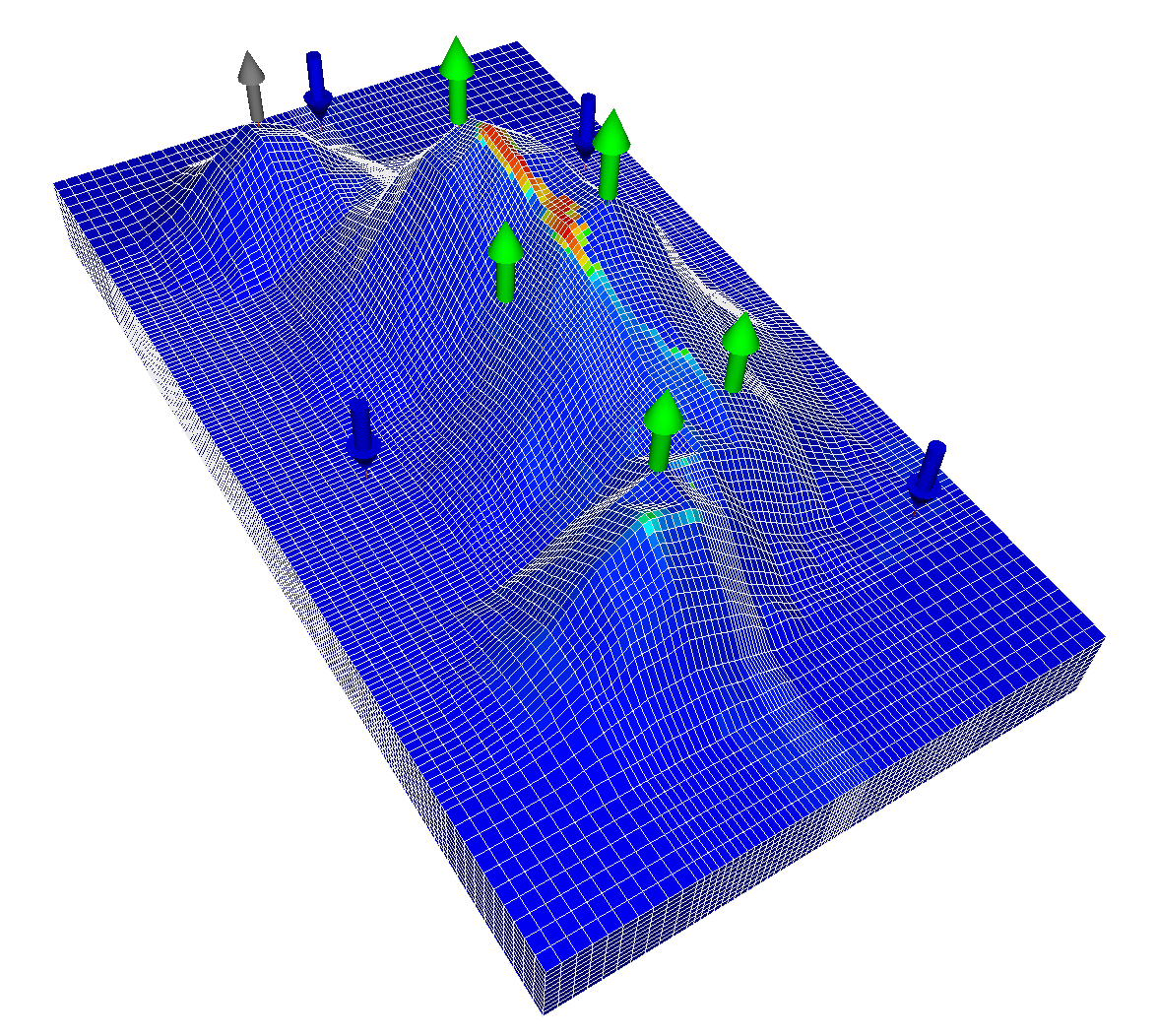}
    \caption{Gas saturation map for the anticlinal model. The water injectors are the blue arrows and the producers are the green arrows. Plotted with ResInsight \cite{ResInsight}.}
    \label{ANTICLINE_BO}
\end{figure}

\begin{figure}
    \centering
    \includegraphics[width=0.7\textwidth]{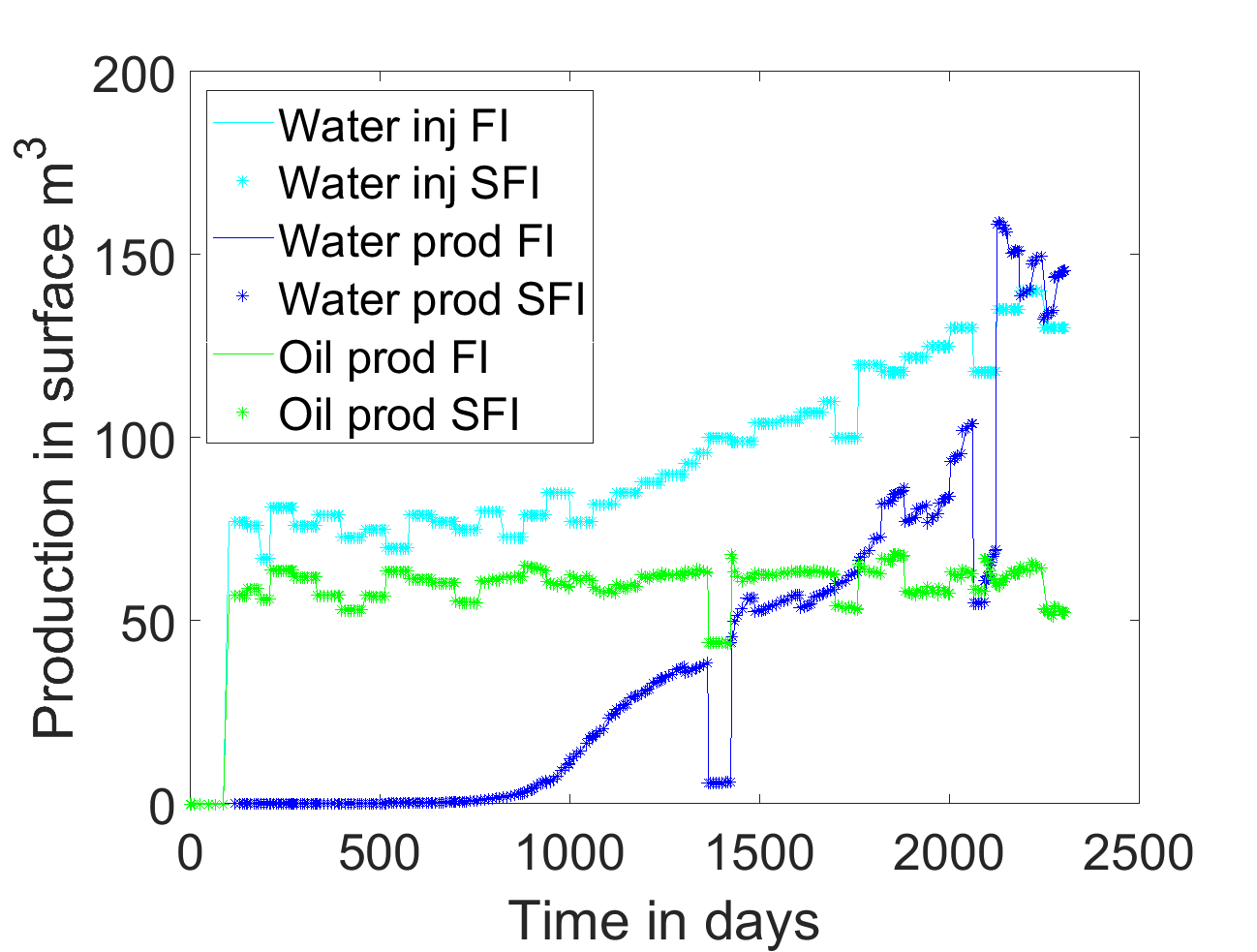}
    \caption{Liquid rates (water injection/production and oil production) profiles for the FI and SFI methods for the anticlinal model.}
    \label{ANTICLINE_BO_FLR}
\end{figure}

\begin{figure}
    \centering
    \includegraphics[width=0.7\textwidth]{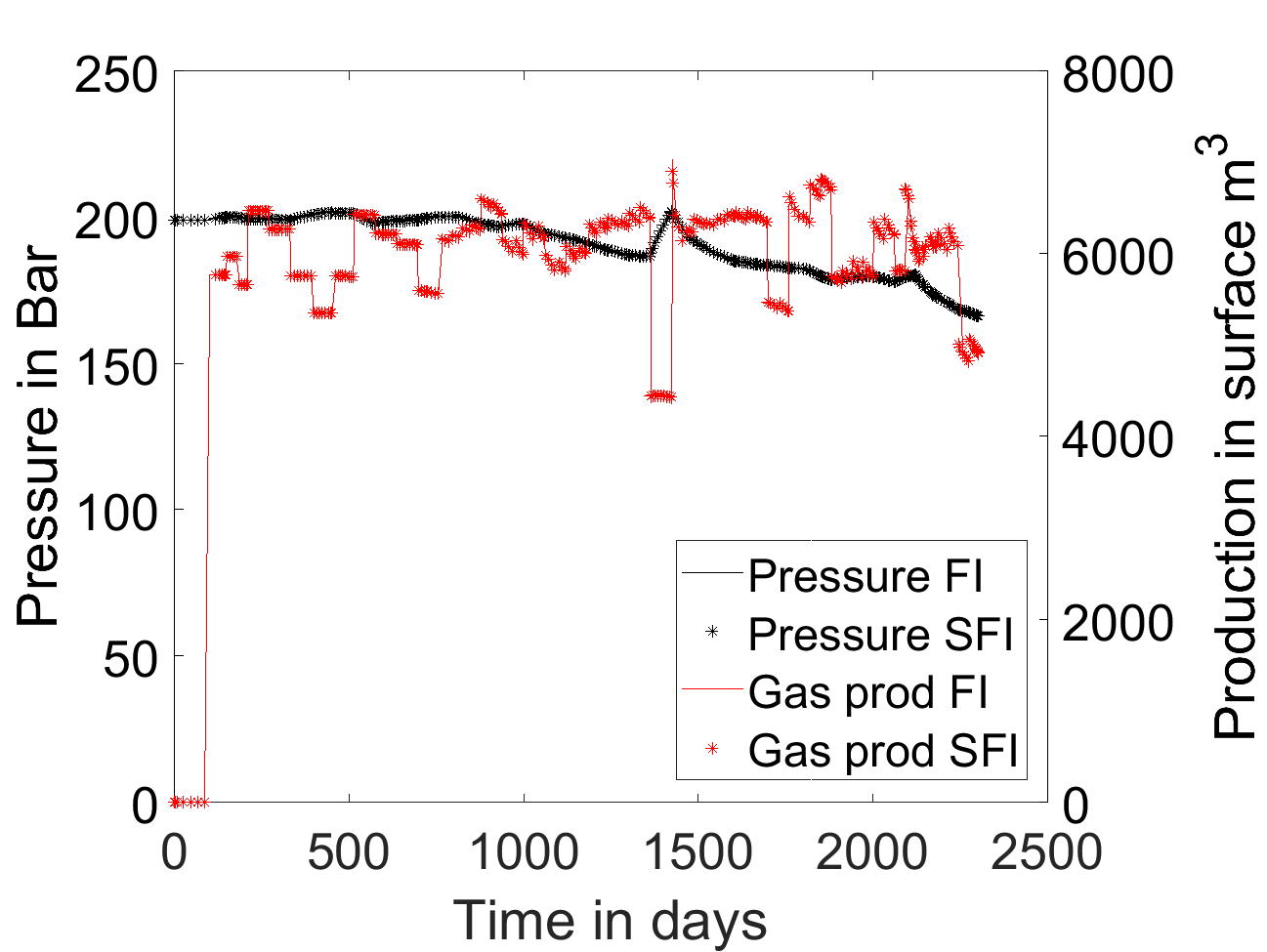}
    \caption{Average reservoir pressure and gas production rate profiles for the FI and SFI methods for the anticlinal model.}
    \label{ANTICLINE_BO_FPR}
\end{figure}

\begin{figure}
    \centering
    \includegraphics[width=0.7\textwidth]{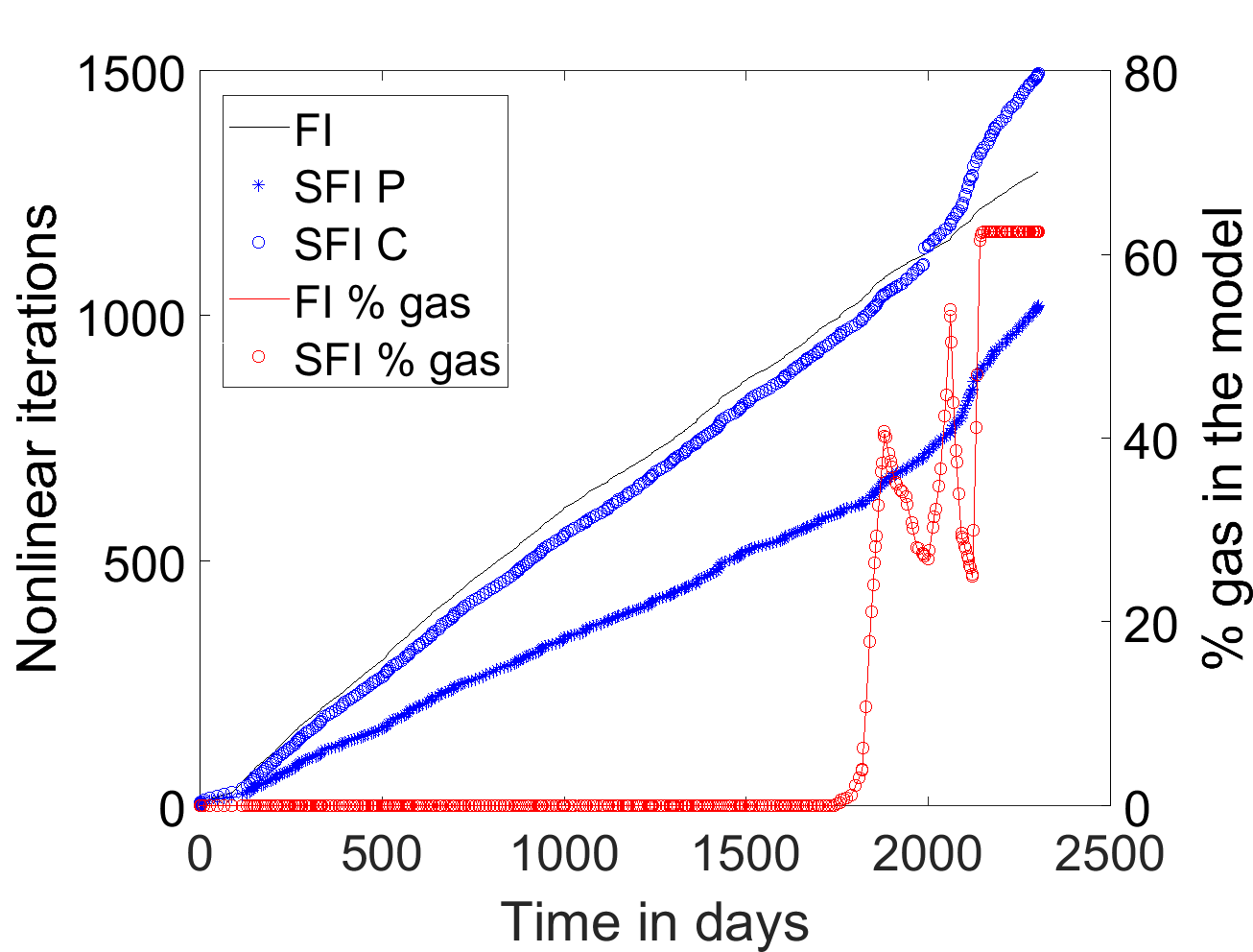}
    \caption{Cumulative Newton iterations for the FI method and cumulative pressure and composition iterations for the SFI method for the anticlinal model. The percentage of gas in the model for FI and SFI methods are also shown.}
    \label{ANTICLINE_BO_ITER}
\end{figure}

\begin{figure}
    \centering
    \includegraphics[width=0.7\textwidth]{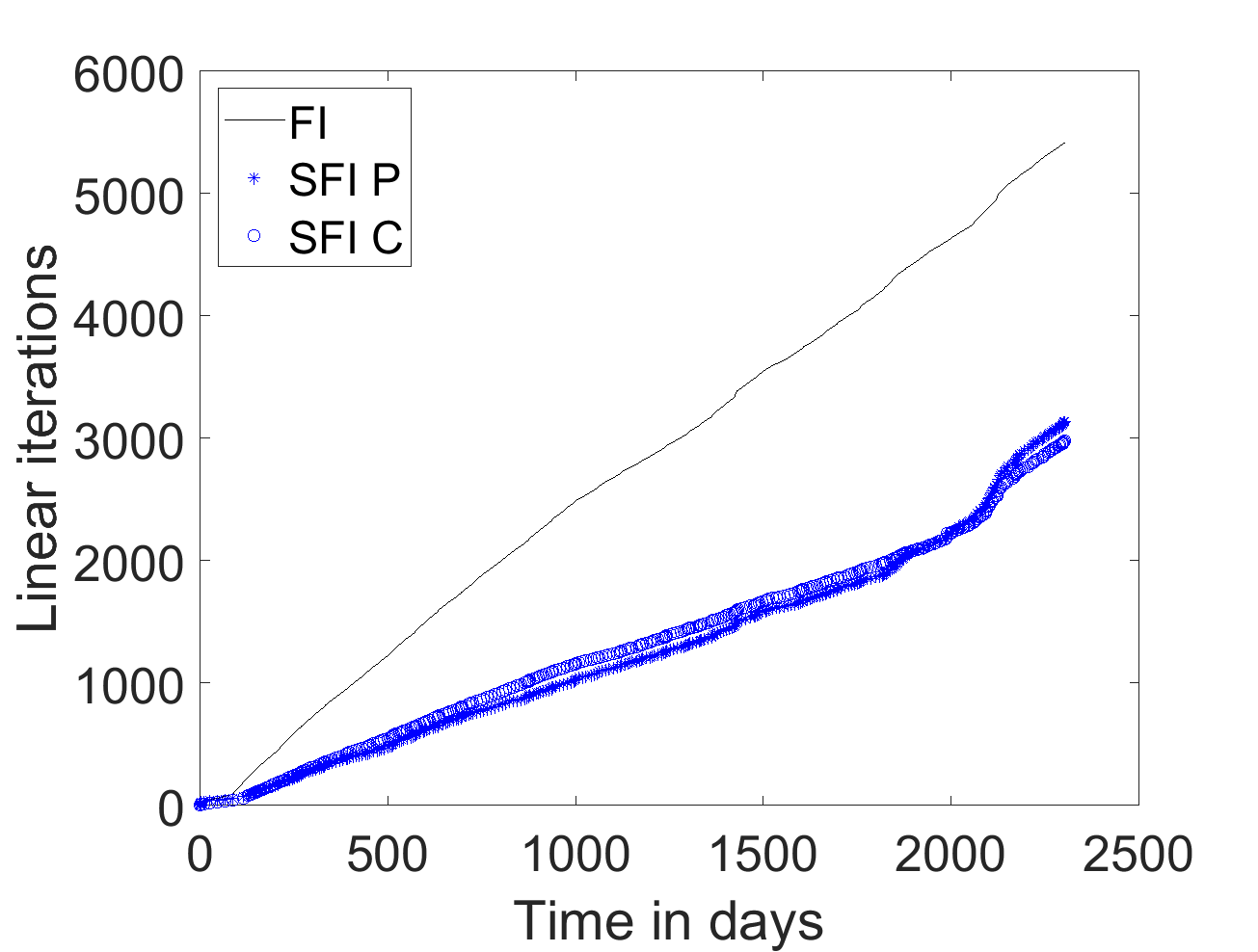}
    \caption{Cumulative CPR iterations for the FI method and cumulative linear iterations for the pressure and composition systems for the SFI method for the anticlinal model.}
    \label{ANTICLINE_BO_LINS}
\end{figure}

The 3D model here represents an anticlinal formation.
The fluid is modeled as a two-component compositional black-oil formulation. There are four water injectors and six producers.
Figure \ref{ANTICLINE_BO} shows the model and the gas saturation at the end of the simulation.
Figures \ref{ANTICLINE_BO_FLR} and \ref{ANTICLINE_BO_FPR} show the injection and production profiles for each phase as well as the average reservoir pressure.
We observe that the FI and SFI produce exactly the same results. 
The FI method converges in 304 timesteps and the SFI method in 361 timesteps.
Figure \ref{ANTICLINE_BO_ITER} shows the cumulative Newton iterations for the FI method, the cumulative pressure and composition iterations for the SFI method as well as the percentage of gas in the model for both the FI and SFI methods.
We see that the percentage of gas in the model changes quite rapidly, and that both methods capture exactly the same physics.
These rapid changes are due to the complexity of the medium with strong contrasts between the geological facies.
The SFI method requires 20\% fewer pressure iterations than the FI method requires Newton iterations and 
15\% more composition iterations than the FI method requires Newton iterations.
Figure \ref{ANTICLINE_BO_LINS}, shows the cumulative number of CPR(AMG,ILU0)-FGMRES iterations for the FI method and
the pressure solves of the SFI method and the cumulative number of ILU0-GMRES iterations for the compositional system of the SFI method.
The CPR(AMG,ILU0)-FGMRES system of the FI method requires in average 4.2 iterations,
while the pressure system of the SFI method requires in average 3.1 linear iterations and the
composition system of the SFI method requires in average 2 linear iterations per system.
As a result, the number of linear iterations from the pressure systems of the SFI method amount to 58\% of the number of CPR iterations for the coupled system of the FI method,
and the number of linear iterations from the composition systems of the SFI method amount to 55\% of the number of CPR iterations for the coupled system of the FI method.

% ANTICLINE_BO_FIM_TRANSBELLP8C8_TS20_QT40_CFLSEQ100

%\bigskip
%\noindent

%% main text
\section{Conclusions}

We proposed a new compositional Sequential Fully Implicit (SFI) solution scheme with a nonlinear pressure equation for compositional flow simulation. 
The compositional SFI scheme is an extended natural-variables formulation.
The pressure equation used here is obtained as a weighted sum of the nonlinear component conservation equations, and it can be seen as a nonlinear overall-volume balance. 
This pressure equation is different from the `usual' pressure equations used in sequential formulations, which are obtained as a linear combination of the linearized component 
conservation equations or by linear combination of the component conservation equations with constant coefficients.
The nonlinear volume-balance equation simplifies to the usual volume-balance equations used by SFI methods for dead-oil and black-oil systems.
There are two splitting errors associated with the compositional SFI scheme: (1) volume balance errors, and (2) total-velocity errors. Our analysis indicates that the volume-balance errors
are highly localized around fronts where a phase change takes place; however, the errors in the total-velocity field can be quite large with large spatial support that spans the
invaded regions of the reservoir.
The  total-velocity error associated with the sequential-implicit split is the primary cause of nonlinear convergence problems
for SFI schemes including ours. Here, we demonstrate that the splitting errors can be reduced to arbitrarily small values when a nonlinear pressure equation is used. Moreover, we show how to effectively 
control the splitting errors in order to achieve convergence to any desired tolerance. 
We tested the compositional SFI approach for challenging test cases in 1D, 2D and 3D.
For strongly coupled problems, the SFI algorithm requires a few more iterations than the FI method, but for weakly coupled cases, 
the sequential algorithm requires fewer iterations than the FI method. Overall, the compositional SFI scheme enjoys convergence properties that are very competitive with the fully-implicit method,
even when the coupling between the flow and the transport is quite strong. 
The SFI formulation has several advantages, however. With the sequential formulation, the parabolic and hyperbolic operators of compositional multi-component flow and transport are now 
identified and decoupled.
As a result, it is now possible to design specific numerical methods optimized for each sub-problem. That includes the use of specific linear solvers for the parabolic and hyperbolic operators, as well as 
opening the door for advanced (high-order) spatial and temporal discretization schemes. 
Finally, this new SFI scheme is well suited for multiscale compositional formulations that rely on sequential coupling of the flow and transport problems.

%% main text
\newpage 
\section{Nomenclature}
\begin{itemize}
   \item $\phi$ porosity of the rock
   \item $K$ permeability in mD
   \item $D$ depth in ft
   \item $g$ gravitational acceleration in ft$^2$-Psi/lb
   \item $n_h$ number of hydrocarbon components
   \item $MW_c$ molar fraction of component~$c$
   \item $P_p$ pressure of phase $p$ in Psi
   \item $P_{c_{GO}}(S_g)=P_g-P_o$ capillary pressure in Psi
   \item $P_{c_{OW}}(S_w)=P_o-P_w$ capillary pressure in Psi
   \item $S_p$ flow saturation of phase~$p$
   \item $S^{thermo}_p$ thermodynamic saturation of phase~$p$
   \item $\overline{S}^{thermo}_p$ normalized thermodynamic saturation of phase~$p$
   \item $x_c$ mole-fraction of component~$c$ in the oil phase
   \item $y_c$ mole-fraction of component~$c$ in the gas phase
   \item $z_{\alpha}$ total mole-fraction of component~$\alpha=c,w$
   \item $\beta_p$ mole-fraction of phase~$p$
   \item $\rho_p$ phase mole density in lbmol/ft$^3$
   \item $\rho_t$ total mole density in lbmol/ft$^3$
   \item $\overline{\rho}_p$ phase mass density in lb/ft$^3$, $\overline{\rho}_g = (\sum_c y_c MW_c) \rho_g$, $\overline{\rho}_o = (\sum_c x_c MW_c) \rho_o$ and $\overline{\rho}_w = MW_w \rho_w$.
   \item $\mu_p$ phase dynamic viscosity in cP
   \item $k_{r_p}$ relative permeability for phase $p$
   \item $\hat{f}_{c,p}$ fugacity of hydrocarbon component~$c$ in phase $p$ in Psi
   \item $u_p$ velocity of phase $p$ in ft/day
   \item $u_t$ total-velocity in ft/day
   \item $u^f_t$ frozen total-velocity during the composition update in ft/day
   \item $\lambda_p$ mobility for phase $p$ in cP$^{-1}$
   \item $\lambda_t$ total mobility in cP$^{-1}$
   \item $q_p$ and $q_t$ well contributions (source or sink) in lbmol/day
   \item $R_{thermo}$ thermodynamic volume splitting error
   \item $R_{u_t}$ normalized total-velocity splitting error
   \item $R_{\nabla \cdot  u_t}$ normalized divergence of total-velocity splitting error
   \item $\overline{\nabla \cdot  u_t}$ dimensionless divergence of the total-velocity 
   \item $N_{\alpha}$ number of moles of component~$\alpha=c,w$ in lbmol
   \item $V_T$ total fluid volume in ft$^3$
   \item $V_P$ discretized pore volume in ft$^3$
   \item $Q_t$ total volumetric rate in ft$^3$/day
   \item $c_t$ total compressibility in Psi$^{-1}$
   \item $V_{T_{\alpha}}$ partial molar volume of component~$\alpha=c,w$ in ft$^3$/lbmol
   \item $\xi_{\alpha}$ density of component~$\alpha=c,w$ in lbmol/ft$^3$
\end{itemize}

%% main text

\section{Acknowledgments}

The authors would like to thank TOTAL management for permission to publish this work and O. M{\o}yner for constructive discussions.

%\appendix

\section{Appendix A: Description of the Models}

% Grid info
The 1D model used in this study has 220 cells of dimensions of $dx=20$ ft, $dy=10$ ft, $dz=2$ ft. 
The porosity is $\phi=0.2$ and the permeability $K=100$ mD.
The 2D model has 220$\times$60 cells with porosity and permeabilities from the top layer of SPE~10 comparative solution project \cite{SPE10}. 
% SCAL/rock
The rock compressibility for the 1D and 2D cases is 1.78e-5 1/Psi. 
The relative permeabilities are quadratic for both the oil and the gas phases. 
The capillary pressures are zero.
% PVT
The fluid is taken from the SPE~5 comparative solution project \cite{SPE5}.
% Initial state/composition
The test cases have initial conditions with $P=4000.0$ Psi, $T=160^{\circ}$F and $S_o=1.0$; table \ref{SPE5_INIT} provides the initial composition values.
\begin{table}[h]
\scriptsize
\centering
\caption{Initial oil composition for SPE~5 fluid.}
\label{SPE5_INIT}
\begin{tabular}{|c|c|c|c|c|c|}
\hline
$C_1$ & $C_2$ & $C_3$ & $C_4$ & $C_5$ & $C_6$ \\
\hline
0.5 & 0.03 & 0.07 & 0.2 & 0.15 & 0.05 \\
\hline
\end{tabular}
\end{table}
%
% Well controls
The connection factors for all the cell-well connections are fixed at $0.3$~Rbbl.cP/day-Psi.
The fluid is produced at 1800 Psi during the depletion.

\section{Appendix B: SI Metric Conversion Factors}
\begin{tabular}{|c|c|c|c|c|}
 \hline
 1 & Psi & = &  100000/14.5037 & Pa \\
 1 & ft & = & 1/3.2808  &  m \\
 1 & lb & = & 1/2.20462262 & kg \\
 1 & mD & =  & 9.869e-16 &  m$^2$ \\
 1 & cP & = & 0.001  &  Pa.s \\
 1 & day & = & 86400 & s \\
 \hline
\end{tabular}

%\appendix

% compile 1 time latex, 1 time bibtex and 2 times latex
\section{References}
\bibliographystyle{plainnat}
\bibliography{bibliography}

\end{document}